\documentclass[12pt]{article}
\pdfoutput=1
\usepackage{tikz}
\usetikzlibrary{matrix,arrows,positioning,shapes,snakes,fadings,decorations.pathmorphing,
decorations.pathreplacing,automata,shadows,decorations.markings,patterns}
\usetikzlibrary{calc,shapes.callouts,shapes.arrows}
\usepackage{jheppub}
\usepackage{amsmath,amssymb,euscript,array,mathrsfs,appendix,ctable,marvosym}
 
\newcommand \arXivlink [1]{\href{http://arxiv.org/abs/#1}{\tt arXiv:#1}}

\usepackage{titlesec}
\titleformat{\subsection}[display]{\it}{}{0.1cm}{\vspace{-1.5cm}\begin{center}\thesubsection\hspace{0.2cm}}[\end{center}\vspace{-0.5cm}]

\newcommand{\ket}[1]{|#1\rangle}
\newcommand{\tr}{\text{tr}}

\newcommand{\EQ}[1]{\begin{equation}\begin{split} #1
\end{split}\end{equation}}

\title{Hawking Radiation Correlations of Evaporating Black Holes in JT Gravity}
\author[a]{Timothy J. Hollowood,}
\author[a]{S.~Prem Kumar}
\author[b]{and Andrea Legramandi}
\affiliation[a]{Department of Physics, Swansea University, Swansea, SA2 8PP, U.K.}
\affiliation[b]{Dipartimento di Fisica, Universit\`a di Milano–Bicocca, Piazza della Scienza 3, I-20126 Milano, Italy}
\emailAdd{t.hollowood@swansea.ac.uk, s.p.kumar@swansea.ac.uk, a.legramandi@campus.unimib.it}
\abstract{
We consider the Hawking radiation emitted by an evaporating black hole in JT gravity and compute the entropy of arbitrary subsets of the radiation in the slow evaporation limit, and find a zoo of possible island saddles. The Hawking radiation is shown to have  long range correlations. We compute the mutual information between early and late modes and bound from below their squashed entanglement. A small subset of late modes are shown to be correlated with modes in a suitably large subset of the radiation previously emitted as well as later modes. We show how there is a breakdown of the semi-classical approximation in the form of a violation of the Araki-Lieb triangle entropy inequality, if the interior of the black hole and the radiation are considered to be separate systems. Finally, we consider how much of the radiation must be collected, and how early, to recover information thrown into the black hole as it evaporates.
}

\notoc
\begin{document}

\maketitle

\newpage

\section{Introduction}

Recent work \cite{Penington:2019kki,Almheiri:2019qdq} has led to a step change in understanding the information loss paradox of black holes. We can now see the missing ingredient in Hawking's calculation \cite{Hawking:1976ra} using only semi-classical methods: there are new saddle points of the functional integral used for calculating entropies of quantum fields on the black hole geometry, the replica wormholes. These new saddles allow one to calculate semi-classically the Page curve \cite{Page:1993wv} of a black hole. A pure state of matter that collapses to form a black hole will return to a pure state. The replica wormholes, along with Page's original insight, imply that Hawking radiation must have very non-trivial correlations between subsets emitted at different times. It is the purpose of this paper to probe these correlations for an evaporating black hole in the JT gravity set up for which the replica wormhole calculations have fully developed.

Quantum information tools have provided a very powerful way to understand the behaviour of black holes. This goes back to tracking the entropy of the Hawking radiation across the lifetime of an evaporating black hole, the Page curve  \cite{Page:1993wv}, or the way that the information of systems, e.g.~a diary, dropped into a black hole, can be recovered in the radiation, considered by Hayden and Preskill   \cite{Hayden:2007cs}. The Page curve has been derived in these, or related, scenarios \cite{Penington:2019kki,Almheiri:2019psf,Penington:2019npb,Almheiri:2019yqk,Almheiri:2019qdq} (see also \cite{Gautason:2020tmk,Hashimoto:2020cas,Hollowood:2020cou}). It is a goal of this work, to show that replica wormhole techniques, and the effective rules that they give rise to, mean that more refined information processing properties of black holes can also be calculated from first principles via standard quantum field theory calculations. For instance, we are able to derive the detailed aspects of information recovery anticipated by Hayden and Preskill. We will show that the information contained in a diary, in this case taking the form of a local quench in the QFT, thrown into a black hole before the Page time, can be recovered at a time
\EQ{
t=t_\text{Page}+\frac1k\cdot\frac{S_\text{diary}}{S_\text{BH}}\ ,\qquad t_\text{Page}=\frac 2k\log\frac32\ ,
}
where $S_\text{diary}$ is the entropy of the diary assumed to be small compared with $S_\text{BH}$, the Bekenstein-Hawking entropy of the black hole, and $k$ is the evaporation rate.

It has been appreciated for a while, that if black holes obey the laws of quantum mechanics then something quite dramatic must happen to reconcile unitarity of evaporation with the rules of effective theories. More precisely, if a Hawking mode $B$ is emitted by an old black hole, one past the Page time, then it must be entangled with a mode of the early Hawking radiation $R_B$ in order to ensure unitarity. On the other hand, the usual rules of local effective theory imply that the Hawking mode must be entangled with its partner mode behind the horizon $A$. Quantum mechanics, of course, forbids $B$ to be maximally entangled with two separate subsystems. One reaction is to give up the entanglement across the horizon leading to a separable quantum state with a diverging energy density at the horizon, a ``firewall" \cite{Almheiri:2012rt}. Another, arguably even more dramatic answer that maintains a smooth geometry at the horizon, is to hypothesize that $A$ and $R_B$ are not separate subsystems $A{=}R_B$: modes on the inside of the black hole are actually living in the Hilbert space of the early Hawking radiation (see \cite{Harlow:2014yka}). 

One of the goals of this work is to pin down where $R_B$ lies within the early radiation by taking $B$ to be a small subset of modes emitted at a certain time by an old black hole. We then attempt to locate $R_B$ by maximizing $B$'s mutual information with the radiation emitted earlier.\footnote{This is a necessary condition because the mutual information by itself does not imply entanglement. However, we also consider a genuine measure of entanglement known as the {\it squashed entanglement\/}. Half the mutual information is an upper bound for the latter.} We find that $R_B$ must lie in a large subset of modes emitted from around the Page time to just before $B$. This means that the purifier $R_B$ is de-localized in the earlier radiation. This is in tune with ideas from from quantum information theory that suggest that extracting $R_B$ would be computationally a hard problem \cite{Harlow:2013tf}.

The astonishing ``$A{=}R_B$" scenario grew out of ideas of {\it black hole complementarity\/} \cite{Susskind:1993mu,Bousso:2012as,Banks:2012nn,Papadodimas:2012aq,Verlinde:2012cy,Maldacena:2013xja,Harlow:2013tf,Papadodimas:2013wnh,Papadodimas:2013jku,Verlinde:2013qya} and ER=EPR \cite{Maldacena:2013xja} (see the review \cite{Harlow:2014yka} for a detailed discussion and other references). If $A{=}R_B$ is really true, then it is legitimate to ask what goes wrong with conventional effective quantum field theory on the black hole background? It is a goal of this work to show that one way that the breakdown of QFT manifests as a breakdown of the consistency conditions on the entropies of spatially separated regions of the quantum fields, specifically the triangle inequality of Araki and Lieb \cite{Araki:1970ba} 
\EQ{
S_{AR}\geq \big|S_A-S_R\big|\ .
\label{tie}
}
Here, $A$ will be modes behind the horizon and $R$ the Hawking modes of an old black hole. The reason for the breakdown will be that for an old black hole, $S_R$ is dominated by a replica wormhole saddle (has an ``island") whereas $S_A$ and $S_{AR}$ are not. It is the island then that disrupts the usual consistency of the entropies in QFT. In retrospect, the conclusion is not surprising because the triangle inequality does not apply to subregions that overlap, and when $R$ has an island then this overlaps with $A$.

The organization of this paper is as follows. In section \ref{s2}, we review the set up in JT gravity, where an evaporating black hole is created by a local quench \cite{Hollowood:2020cou} (related to scenarios in \cite{Engelsoy:2016xyb,Almheiri:2019psf}). In this section, we emphasize the simplifications that occur in the slow evaporation limit. In section \ref{s3}, we describe how to evaluate the entropy of a set of intervals in the bath, including the island saddles that follow from the replica wormholes. The key computation is the solution for the island saddles that we show simplifies in the slow evaporation limit. There are a whole zoo of island saddles, even with the assumptions we make, but usually only a few are actually needed. The remaining sections put our entropy formulae to use. In section \ref{s4}, we derive the Page curve of the evaporating black hole and extract the Page time. We then calculate the correlation between the early and late Hawking radiation in the form of the mutual information. This shows that there are strong correlations, as expected on the basis of Page's analysis. The correlations can be shown to be quantum, i.e.~entanglement, by establishing a lower bound on the {\it squashed entanglement\/}, a measure of entanglement in mixed states. 
In section \ref{s5} we analyse the correlation between the early and late Hawking modes in more detail. We pick a narrow interval of late modes $B$ and find out which interval of early modes it is maximally correlated with. This establishes that modes entangled with $B$, $R_B$, are de-localized over a large subset of the early modes that extends from around the Page time to the modes emitted just before $B$. This is what is expected: the entangled modes $R_B$ should be difficult to extract from the early radiation \cite{Harlow:2013tf}. Section \ref{s6} is devoted to showing that, when the interior of the black hole is considered, there is breakdown of the Araki-Lieb triangle inequality for the entropies of the interior and the radiation. This provides a smoking gun for the $A{=}R_B$ scenario. In section \ref{s7} we consider how information thrown into the black hole in the form of an entropy carrying local quench in the CFT is recovered in the Hawking radiation. We find detailed agreement with the quantum information analysis of Hayden and Preskill \cite{Hayden:2007cs}. Finally in section \ref{s8} we show an operator insertion behind the horizon is observable in the bath, if the appropriate interval in the bath is in its island saddle.

\subsection{Entropy as an observable}\label{s1.1}

For the new developments involving black holes and the information loss paradox, the entropy plays a key r\^ole. Our results involve the entropies of sub-regions in the radiation bath and associated quantum information measures. It is a natural question to ask whether these entropies are observable, even in principle? This is important because we have shown that the entropy of sub-regions of the bath are sensitive to physics behind the horizon and so it is fundamental to understand if it is actually observable from the bath.

More generally, we can consider the R\'enyi entropies $S_A^{(n)}=(1-n)^{-1}\log\text{tr}\,\rho_A^n$ of the subregion, where $S_A=\lim_{n\to1}S_A^{(n)}$. Note that for a finite dimensional subsystem $A$ of dimension $d_A$, only the first $d_A-1$ R\'enyi entropies are needed to extract the eigenvalues of $\rho_A$ and hence the von Neumann entropy of $A$.

The R\'enyi entropies are not directly conventional observables in the sense of being associated to a Hermitian operator. They can be computed by joint measurements on $n$ copies of the system. However, at least in the case of finite dimensional systems, they can be measured by measuring a set of conventional, but random, observables on a single copy of the system (e.g.~\cite{EB,EVRZ}). The idea is take a complete set of rank-1 projection operators on $A$, $\Pi_j$, $j=1,2,\ldots,d_A$. Then define the rotated sets $U\Pi_jU^\dagger$, for an arbitrary unitary operator $U$ on $A$. These are associated to some Hermitian operators ${\cal O}_U=\sum_j\lambda_jU\Pi_jU^\dagger$. Then one measures ${\cal O}_U$ in the conventional sense on copies of the system in order to estimate the Born rule probabilities for the outcomes $j$ and for an arbitrary $U$,
\EQ{
p_U(j)=\text{tr}\big(\rho_AU\Pi_jU^\dagger\big)\ .
}
The estimates for the R\'enyi entropies are then given as averages of powers of the probabilities in the unitary ensemble; for example, for the second R\'enyi entropy
\EQ{
\text{tr}\,\rho_A^2=(d_A+1)\sum_j\overline{p_U(j)^2}-1\ ,
}
where the over-line indicates an average over the unitary orientation. The higher $\text{tr}\,\rho_A^n$ involve a similar average of a polynomial of order $n$ in the probability. The explicit formula is given in \cite{EVRZ}:
\EQ{
\sum_j\overline{p_U(j)^n}=\sum C_{b_1,\ldots,b_n}\prod_{j=1}^n\big(\tr\rho_A^j\big)^{b_j}\ ,
}
where the sum is over conjugacy classes $1^{b_1}2^{b_2}\cdots n^{b_n}$ of the symmetric group $S_n$ and
\EQ{
C_{b_1,\ldots,b_n}=\frac{n!}{\prod_{j=1}^n(d_A+j-1)j^{b_j}b_j!}\ .
}
In a real application of this protocol, the average over the unitary ensemble is realized in terms of a discrete sampling known as a $k$-{\it unitary design\/}.

The conclusion is that, in principle, the entropy can be measured locally on a subsystem using conventional, albeit random, quantum measurements, at least for finite dimensional systems.   

\section{The evaporating black hole}\label{s2}

The setup consists of the extremal black hole in Jackiw-Teitelboim gravity \cite{Jackiw:1984je,Teitelboim:1983ux} defined on a patch AdS$_2$, with the standard metric in Poincar\'e coordinates 
\EQ{
ds^2=-\frac{4dx^+dx^-}{(x^--x^+)^2}\ ,
}
with a half Minkowski space spliced on the time-like boundary to act as a radiation bath with metric $ds^2=-dy^+dy^-$ \cite{Engelsoy:2016xyb,Almheiri:2019psf,Almheiri:2019yqk,Almheiri:2019qdq, Almheiri:2020cfm}. Then along the boundary $x^+\sim x^-$,\footnote{The details of the regularization at the boundary are described in \cite{Engelsoy:2016xyb}.} we have $x^\pm=y^\pm$. The boundary conditions are transparent, so that modes of the CFT propagate through the boundary without reflection. On the AdS region, the gravitational sector includes the dilaton, which in the extremal black hole case takes the form
\EQ{
\phi=\phi_0+\frac{2\phi_r}{x^--x^+}\ .
}
A CFT, which we take to be a large number of free fermions, propagates across the whole geometry, the AdS and Minkowski regions. 

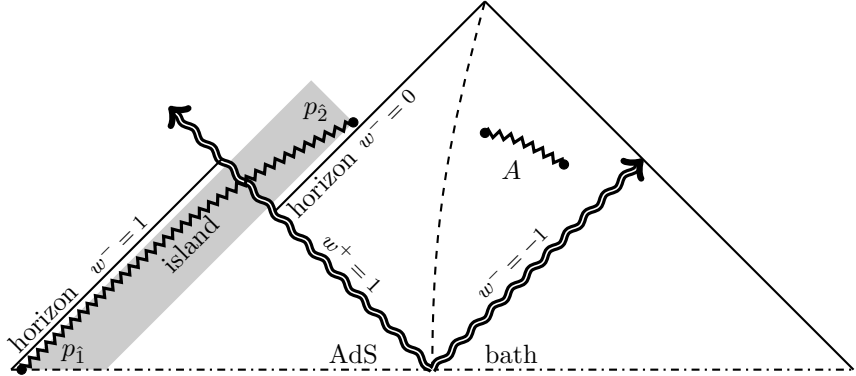
\begin{figure}[ht]
\begin{center}
\begin{tikzpicture} [scale=0.7]
\filldraw[color=black!20,draw=none] (-7.8,0) -- (-6.2,0) -- (-1.5,4.7) -- (-2.3,5.5) -- (-7.8,0); 
\draw[thick,dashed] (0,0) to[out=90,in=-105] (1,7);
\draw[thick] (1,7) -- (8,0);
\draw[thick,dash dot] (-8,0) -- (8,0);
\draw[double,decorate,very thick,->,decoration={snake,amplitude=0.03cm}] (0,0) -- (-5,5);
\draw[double,decorate,very thick,->,decoration={snake,amplitude=0.03cm}] (0,0) -- (4,4);
\draw[thick] (-8,0) -- (-4,4);
\draw[thick] (-3,3) -- (1,7);
\draw[decorate,very thick,decoration={zigzag,segment length=1.5mm,amplitude=0.5mm}] (1,4.5) to[out=-5,in=150] (2.5,3.9);
\draw[decorate,very thick,decoration={zigzag,segment length=1.5mm,amplitude=0.5mm}] (-7.8,0) to[out=43,in=-155] (-1.5,4.7);
\filldraw[black] (1,4.5) circle (2.5pt);
\filldraw[black] (2.5,3.9) circle (2.5pt);
%
%
\filldraw[black] (-7.8,0) circle (2.5pt);
\filldraw[black] (-1.5,4.7) circle (2.5pt);
\node at (1.5,0.3) {\footnotesize bath};
\node at (-1.5,0.3) {\footnotesize AdS};
\node[rotate=45] at (-2.1,3.55) {\footnotesize horizon};
\node[rotate=45] at (-7.4,1) {\footnotesize horizon};
\node at (-6.8,0.3) {\footnotesize $p_{\hat1}$};
\node at (-2.2,4.9)  {\footnotesize $p_{\hat2}$};
\node at (1.5,3.8) {\footnotesize $A$}; 
\node[rotate=45] at (-4.6,2.4) {\footnotesize island};
\node[rotate=45] at (-0.9,4.8) {\scriptsize $w^-=0$};
\node[rotate=45] at (-6,2.4) {\scriptsize $w^-=1$};
\node[rotate=-45] at (-1.5,2.05) {\scriptsize $w^+=1$};
\node[rotate=45] at (1.5,2.05) {\scriptsize $w^-=-1$};
\end{tikzpicture}
\caption{\footnotesize A Penrose diagram showing a shockwave inserted at the boundary of the AdS and bath regions with an in-going component that excites the extremal black hole, shifting the horizon out, leading to an evaporating black hole. Also shown is an interval $A$ in the bath and its island, the shaded area in the AdS region, the causal domain of the 2 QESs $p_{\hat a}$.}
\label{fig1} 
\end{center}
\end{figure}

The evaporating black hole is created by a local quench \cite{Hollowood:2020cou}---an operator insertion---in the CFT initiated from a point on the boundary at $t=0$ that leads to an in-going and out-going shockwave: see figure \ref{fig1}.\footnote{We choose Penrose diagrams so that the straight line at the bottom corresponds to $t=0$ in the bath and $t_\text{Poincar\'e}=0$ in the AdS region. The boundary between the AdS region and the bath, $x^+=x^-$, or $y^+=y^-$, becomes curved behind the shockwaves.} The quench corresponds to a CFT state created by the action of a local operator on the vacuum at the time-like boundary :
\EQ{
{\cal O}(y^\pm=i\varepsilon)\ket{0}\ ,
\label{lic}
}
where the small  shift $\varepsilon$ in the imaginary time direction is needed to ensure that the state can be normalized. The in-going component of the shockwave, $x^+=0$, carries energy into the black hole and excites it to a black hole of inverse temperature
\EQ{
\beta=\sqrt{\frac{\pi\phi_r}{4G_NE_\text{shock}}}\ ,
}
where $E_\text{shock}=\Delta_{\cal O}/\varepsilon$. When $\varepsilon$ is small, the shockwave energy is large and its energy-momentum tensor becomes concentrated on the two wavefronts $x^+=0$ and $y^-=0$ propagating into, and away from, the black hole, respectively. 

In the following, we will suppose that the intrinsic entropy of the shockwave between the in- and out-going components is vanishing, or at least small compared with the gravitational entropy, and can be ignored. For simplicity, we will also assume that the gravitational entropy of the excited black hole is much greater than the extremal entropy
\EQ{
S_\text{BH}=\frac{\pi c}{6\beta k}\gg S_0=\frac{\phi_0}{4G_N}\ .
} 
The key physical quantity here is the inverse time scale
\EQ{
k=\frac{G_Nc}{3\phi_r}\ ,
\label{sew}
}
which sets the rate of evaporation of the black hole. We shall work in the limit where $k\ll1$ (relative to the AdS scale which is set to 1) with $\beta$ fixed and time scales with $kt$ fixed. This limit is simply one of expediency rather than necessity that means we can avoid having to resort to numerical techniques.

In front of the shockwave, the AdS and bath coordinates are related via $x^\pm=y^\pm$. We continue this through the shockwave by writing $x^\pm=f(y^\pm)$. The exact expression for $f(t)$ \cite{Almheiri:2019psf} involves modified Bessel functions of the 1st and 2nd kind and can be deduced from \eqref{huf} below, but there is a simple expression for $f(t)$ for times that extend to $\mathscr{O}(k^{-1})$,
\EQ{
f(t)=\frac\beta\pi\tanh\Big[\frac{2\pi}{k\beta}{\cal S}(t)\Big]\quad\text{where}\quad
{\cal S}(t)=\big(1-e^{-kt/2}\big)\theta(t)\ .
\label{sac}
}
This approximation will be adequate for our analysis here, but longer time scales of order
$k^{-1}|\log\beta k|$ require the full Bessel functions. 

The coordinates $x^\pm$ do not extend behind the horizon and so it will prove useful to define new coordinates $w^\pm$ related to $x^\pm$ by a M\"obius transformation\footnote{In an equilibrium situation at temperature $\beta^{-1}$, $w^\pm\,=\,\pm \exp(\pm 2\pi y^\pm/\beta)$ where $y^\pm$ cover the black hole patch outside the horizon.}
\EQ{
x^\pm=\pm\frac\beta\pi\cdot\frac{w^\pm\mp 1}{w^\pm\pm1}\ ,
\label{mobius}
}
which cover the region inside the horizon of the newly created black hole, $w^->0$. The 
horizon of the original extremal black hole in front of the shockwave is at $w^-=1$, while, behind the shockwave the horizon is at $w^-=0$. It is important that the vacuum state does not change under a M\"obius, or $\text{SL}(2,\mathbb R)$ transformation.

The new coordinates are related to the bath coordinates by an associated function $w^\pm=\pm\hat f(y^\pm)^{\pm 1}$. The function $\hat f$ is expressed in terms of modified Bessel functions of the 1st and 2nd kind \cite{Almheiri:2019psf},
\EQ{
\hat f(t)=\frac{e^{4\pi/(\beta k)}}\pi\frac{K_0(z)}{I_0(z)}\ ,\qquad z=\frac{2\pi}{\beta k}e^{-kt/2}\ .
\label{huf}
}
However, in the small $k$ regime with $kt$ fixed, it takes a simple form which we write through the shockwave, and for times up to $\mathscr{O}(k^{-1})$, as
\EQ{
\log\hat f(t)=\theta(-t)\log \frac{1+\pi t/\beta}{1-\pi t/\beta}+\frac{4\pi}{\beta k}{\cal S}(t)\ .
\label{gre}
}
The expression for $t<0$ yields $x^\pm=t$ via the M\"obius transform \eqref{mobius}.

In what follows, we will need the small $k$ limit, with $kt$ and $\beta$ fixed, of various quantities that depend on $f$ and $\hat f$ evaluated at various times. We will use the notation $\hat f_i=\hat f(t-\tau_i)$, etc.  The leading behaviour we want is always $\mathscr{O}(k^{-1})$. For example, consider $\log\hat f_i$, whose leading behaviour is
\EQ{
\log \hat f_i\to \frac{4\pi}{\beta k}{\cal S}(t-\tau_i)\ .
}
So at this order, $\log\hat f_i$ switches on abruptly at $t=\tau_i$. Another important example is the quantity $\log(\hat f_j^{-1}-\hat f_i^{-1})$ with, say,  $\tau_i<\tau_j$.  In this case, the contribution at order $k^{-1}$ kicks in when $t>\tau_j$, but independent of $\tau_i$,
\EQ{
\log(\hat f_j^{-1}-\hat f_i^{-1})\to -\frac{4\pi}{\beta k}{\cal S}(t-\tau_j)\ .
}
Similarly $\log\hat f_i'\to\frac{4\pi}{\beta k}{\cal S}(t-\tau_i)$. Finally, we can deal with expressions that depend on $f$ rather than $\hat f$ by effectively replacing $f_i\to 1-2\hat f_i^{-1}$, which follows from the M\"obius transformation that relate the two, in the small $k$ limit.

\section{Entropy of arbitrary intervals}\label{s3}

As the excited black hole state evaporates and relaxes, we want to evaluate the von Neumann entropy in the Hawking radiation collected by a certain number of disjoint  intervals in the bath.

\subsection{No-island saddle}

We consider a union $A \equiv \cup_{j=1}^N A_j$ of $N$ disjoint spatial intervals in the bath
\EQ
 {
 A_j=[b_{2j-1},b_{2j}] \quad {\rm with}\quad  b_j<b_{j+1}\,, \qquad  j=1,2\ldots N,
 }
with Minkowski coordinates  for each endpoint $p_j$: $(y^\pm=t\pm b_j)$.  We will be interested in timescales $t$ of order $k^{-1}$ and locations $\{b_j\}$ which are also ${\mathscr O}(k^{-1})$. 
The entanglement entropy of the union of these intervals, in a theory of $2c$ free fermions in the vacuum state in a non-trivial frame,
\EQ{
ds^2=-\Omega^{-2}d\xi^+\,d\xi^-\,,
}
 is given by, 
\EQ{
S_\text{QFT}=-\frac c6\sum_{i<j}(-1)^{j-i}\log \sigma_{ij}-\frac c6\sum_j\log\Omega_j+NS_\text{UV}\ ,
}
where $\sigma_{ij}=-(\xi^+_i-\xi^+_j)(\xi^-_i-\xi^-_j)$ and $S_\text{UV}=-\frac c6\log\varepsilon_\text{UV}$ contains the UV cut off, while $\Omega_j$ are the values of the conformal factor at each endpoint. The result follows from the expression for the entropy of several intervals of free fermions in the Minkowski vacuum  \cite{Casini:2009vk} augmented with a conformal transformation to the non-trivial frame \cite{Almheiri:2019psf}.
 
In the present case, the in-going modes are in the vacuum state of the bath coordinate $y^+$ whereas the out-going modes, containing the Hawking radiation, are in the  vacuum state of the AdS coordinate $w^-$ (or equivalently $x^-$). Hence, the out-going modes---the Hawking radiation---once they are in the bath are manifestly not in the Minkowski vacuum.
 
The QFT entropy, what we will call the ``no-island saddle", is computed by using the frame $(y^+,w^-)$ and takes the form
\EQ{
S_\text{no island}\equiv S_\text{QFT}=-\frac c6\sum_{i<j}(-1)^{j-i}\log(\hat f_j^{-1}-\hat f_i^{-1})(y^+_j-y^+_i)+\frac c{12}\sum_j\log\frac{\hat f_j^2}{\hat f'_j}+\cdots
}
The second set of terms arise from the conformal factors at $\{p_j\}$ from changing coordinates from $y^-$ to $w^-$. Now we take the small $k$ limit, applying the rules that we established earlier. Note that the $y^+$ coordinates do not contribute at leading order in $k^{-1}$:
\EQ{
S_\text{no island}=2S_\text{BH}\sum_{j=1}^{2N}(-1)^{j-1}{\cal S}_{b_j}+NS_\text{UV}\ ,
\label{buk}
}
where we used the shorthand ${\cal S}_b\equiv{\cal S}(t-b)$.
So in the slow evaporation limit, the entropy takes a simple form in terms of the elementary function ${\cal S}(t)$ defined in \eqref{sac}.  The final form above  is consistent with causality. The wavefront of the Hawking radiation as it enters the bath, is along $y^-=0$. It only reaches an endpoint $b$ of an interval at $t=b$ which is reflected in the fact that ${\cal S}(t-b)$ is proportional to the step function $\theta(t-b)$.

\subsection{Island saddles}

The entropy of regions $A=\cup_jA_j$ in the bath that we have just established, can be computed semi-classically using steepest-descent functional integral techniques via the replica method \cite{Calabrese:2004eu,Calabrese:2005in}. The recent step forward, is the realization that in the gravitational context there can be new saddles that involve non-trivial replica geometries, the {\it replica wormholes\/} \cite{Penington:2019kki,Almheiri:2019qdq}. The new saddles, or instantons, can be computed by a recipe as follows: if we are computing an entropy of a region in the bath $A$, then the replica-wormhole saddles require us to define a generalized entropy by adding to the QFT entropy, the contribution of new regions $I$, the island(s), whose endpoints $p_{\hat a}$ are the {\it quantum extremal surfaces\/},\footnote{Strictly speaking the island is the causal domain of a Cauchy slice joining the QES. Note that QESs are just points in 2d gravity, but in higher dimensions they are genuine surfaces.} and also the value of the dilaton at the QESs. Then one extremizes over the positions of the QESs:
\EQ{
S_I(A)=\,\underset{\{p_{\hat a}\}}{\text{ext}}\Big\{\sum_{\hat a}\frac{\phi(p_{\hat a})}{4G_N}+S_\text{QFT}(A\cup I)\Big\}\ .
\label{fuc}
}
Finally, it is the saddle with the smallest entropy amongst the saddles that dominates,
\EQ{
S_A=\,\underset{I}{\text{min}}\,S_I(A)\ ,
}
including the no-island saddle. The recipe of this kind of form first appeared in holographic contexts \cite{Ryu:2006bv,Hubeny:2007xt,Faulkner:2013ana,Engelhardt:2014gca} but the replica wormholes mean that the result can be derived from conventional field theory methods without any recourse to holography. Note that usually, the dilaton terms cost a lot of entropy and so it is only in very special situations that an island saddle dominates over the no-island saddle.
 
 \subsection{Two-QES saddles: one behind shockwave}
 
There are various kinds of possible island saddles that could contribute. We will assume here that the only important ones have 2 QES $p_{\hat a}$, $\hat a=1,2$. Consider the generalized entropy---the terms in the curly brackets in \eqref{fuc}---in $x^\pm$ coordinates
\EQ{
S_\text{gen.}(x_{\hat a}^\pm)&=2S_0+\frac{c}{6}\log(x_{\hat 1}^+-x_{\hat 2}^+)(x_{\hat 2}^--x_{\hat 1}^-)+\frac c6\sum_{\hat a=1}^2\Big\{\frac1 k\cdot\frac1{x_{\hat a}^--x_{\hat a}^+}\\[5pt] &+\log\frac2{x_{\hat a}^--x_{\hat a}^+}
-\sum_{j=1}^{2N}(-1)^{\hat a-j}\log(t+b_j-x_{\hat a}^+)(x_{\hat a}^--f_j)\Big\}+S_\text{no island}\ .
\label{luk2}
}
It is clear from this that extrema would have $x_{\hat a}^\pm=\mathscr{O}(k^{-1})$ and the entropy is dominated at $\mathscr{O}(k^{-1})$ by the no-island contribution. Actually,  closer analysis fails to find a physically acceptable solution and so we will discard this possibility anyway.

Now consider the case with one QES in front and one behind the shockwave, the case illustrated in figure \ref{fig1}. The QES in front $p_{\hat1}$ will have coordinates $x^\pm_{\hat1}=\mathscr{O}(k^{-1})$, i.e.~$w_{\hat1}^\pm=\mp1$, to leading order, and so our task is to find the position of the second QES $p_{\hat2}$ behind the shockwave. 
The ingredients we need, include the conformal factor in the AdS region in the $(y^+,w^-)$ frame,
\EQ{
\Omega^{-2}=\frac{4\hat f'(y^+)}{(1+w^+w^-)^2}\,
\label{xus}
}
and the dilaton,  determined by the master function $\hat f$ \cite{Hollowood:2020cou}:
\EQ{
\phi=\phi_0+2\phi_r\left\{\frac{\hat f''(y^+)}{2\hat f'(y^+)}-\frac{w^-\hat f'(y^+)}{1+w^-\hat f(y^+)}\right\}\ .
}
In the same frame, it is useful to change variable from $y^+$ back to $w^+=\hat f(y^+)$. Within our approximation \eqref{gre}, we then have
\EQ{
\phi&=\phi_0+\frac{2\pi\phi_r}\beta\cdot\frac{1-w^+w^-}{1+w^+w^-}\left(1-\frac{\beta k}{4\pi}\log w^+\right)-\frac{k\phi_r}2\ ,\label{xut}\\[5pt]
\Omega^{-2}&=\frac{4}{(1+w^+w^-)^2}\cdot\frac{2\pi w^+}\beta\left(1-\frac{\beta k}{4\pi}\log w^+\right)\ .
}
Putting the ingredients together, the generalized entropy needed to determine the position of the second QES, is
\EQ{
S_\text{gen.}(w^\pm_{\hat2})&=\frac{\pi c}{6\beta k}\cdot\frac{1-w_{\hat2}^+w_{\hat2}^-}{1+w_{\hat2}^+w_{\hat2}^-}\left(1-\frac{\beta k}{4\pi}\log w_{\hat2}^+\right)\label{box}\\[5pt]
 &+\frac c{12}\log\frac{w_{\hat2}^+(1-\frac{\beta k}{4\pi}\log w_{\hat2}^+)}{(1+w^+_{\hat2}w^-_{\hat2})^2}
+\frac c6\log(w^-_{\hat1}-w^-_{\hat 2})(y^+_{\hat2}-x_{\hat1}^+)\\[5pt]
&-\frac c6\sum_j(-1)^j\log(w_{\hat2}^-+\hat f_j^{-1})(t+b_j-y_{\hat2}^+)+S_\text{no island}+\cdots\ .
}
We have not shown some constants that are sub-leading in $k$.

Extremizing over the coordinates $w_{\hat 2}^\pm$ of the second QES,  leads to a pair of rather complicated equations. However, only a small number of terms actually matter in the small $k$ limit. In particular, the solutions will have $w^+_{\hat2}w^-_{\hat2}={\mathscr O}(k)$ and this allows us to replace $1+w^+_{\hat2}w^-_{\hat2}\to1$. In addition, there are terms contributing to the saddle point equation from varying with respect to $w_{\hat 2}^-$ of the form
\EQ{
\frac{\beta}{2\pi w^+_{\hat2}}\cdot\frac1{y^+_{\hat2}-t-b_j}\ ,\qquad \frac{\beta}{2\pi w^+_{\hat2}}\cdot\frac1{y^+_{\hat2}-x^+_{\hat1}}\ ,}
Since we take $t$, $b_j$ to be   ${\mathscr O}(k^{-1})$, and once we have the solution, $y^+_{\hat2}$ is also ${\mathscr O}(k^{-1})$, these terms are suppressed by a factor of $k$ compared to leading terms in the saddle point equations.

The stripped down equations that determine the leading order behaviour can be written compactly in terms of the variables {$\pmb{ \omega}_{\hat 2}^+$} and $w_{\hat 2}^-$
where 
\EQ{\pmb{ \omega}_{\hat 2}^+=w_{\hat 2}^+\Big(1-\frac{\beta k}{4\pi}\ln w_{\hat 2}^+\Big)=w_{\hat 2}^+ \Big(1-{\cal S}(y_{\hat 2}^+)\Big)\,.
}
It is useful to recall the definition of ${\cal S}$ and note that $\pmb{\omega}_{\hat 2}^+\,=\,w_{\hat 2}^+\,e^{-ky_{\hat 2}^+/2}$. For the saddle point to be behind the shockwave, we need $w_{\hat 2}^+>0$, or equivalently $\pmb{\omega}_{\hat 2}^+>0$. 
Importantly, on the time and length scales of our interest, the pre-factor $(1-{\cal S})$ is ${\mathscr O}(k^0)$. We find,
\EQ{
&\frac{2\pi}{\beta k}{\pmb{\omega}}_{\hat 2}^+-\frac1{w_{\hat 2}^--w_{\hat1}^-} +\sum_{j=1}^{2N}\frac{(-1)^j}{w_{\hat 2}^-+\hat f_j^{-1}}=0\ ,\\
&\frac{2\pi}{\beta k}w_{\hat2}^--\frac1{4\,{\pmb{\omega}}_{\hat2}^+}=0\ .
\label{quf3}
}
The second of these two equations ensures that $w_{\hat 2}^+ w_{\hat 2}^-={\mathscr O}(k)$, consistent with the approximation.

We can now insert the position of the first QES, $w^-_{\hat1}=1$ into the above. If we write $w_{\hat2}^-=1/4p$, then $\pmb{\omega}_{\hat2}^+=\beta kp/2\pi$ and $p$ must satisfy
\EQ{
1+\frac4{4p-1}+4\sum_{j=1}^{2N}\frac{(-1)^j\hat f_j}{4p+\hat f_j}=0\ .
}
The solution of this equation is made straightforward by the extreme behaviour of the functions $\hat f_j$ in the small $k$ limit. At leading order in $k$, it vanishes for $t<b_j$, but then increases exponentially for $t>b_j$ in such a way that $\hat f_j\gg\hat f_i$, if $b_j< b_i$. Hence, one can spot $2N$ distinct solutions labelled by $\alpha=1,2,\ldots,2N$, for which $p=c_\alpha\hat f_\alpha$, i.e.
\EQ{
\pmb{\omega}^+_{\hat2}=\frac{\beta kc_\alpha}{2\pi}\hat f_\alpha\ ,\qquad w^-_{\hat2}=\frac1{4c_\alpha\hat f_\alpha}\ ,
\label{kis}
}
where $c_\alpha=\frac34$, for $\alpha$ odd, and $c_\alpha=\frac1{12}$, for $\alpha$ even.  Working to the leading order in the small $k$ approximation, the coordinate $w_{\hat 2}^+ = \pmb{\omega}_{\hat 2}^+\, e^{ky_{\hat 2}^+/2} $ for the QES behind the shockwave is then simply,
\EQ{
w_{\hat 2}^+\,=\,\frac{\beta kc_\alpha}{2\pi}\hat f_\alpha \,e^{ky_{\alpha}^-/2}\,.
}

We can also write \eqref{kis} in terms of the $w^-$ coordinate of the corresponding point in the bath $w^-_\alpha=-1/\hat f_\alpha$
\EQ{
\pmb{\omega}^+_{\hat2}=-\frac{\beta kc_\alpha}{2\pi}\cdot\frac1{w^-_\alpha}\ ,\qquad w^-_{\hat2}=-\frac{w^-_\alpha}{4c_\alpha}\ ,
\label{kis2}
}
Note that the saddles are only consistent if $w_{\hat2}^+>0$, ensuring that the QES is behind the shockwave: see figure \ref{fig1}. This means that they only appear when $t>b_\alpha$. All these saddles have $w^-_{\hat 2}>0$, and so the second QES is inside the horizon of the black hole created by the shockwave at least in our window of approximation, i.e.~for times $t$ of $\mathscr{O}(k^{-1})$. For much later times, the QES pops outside the horizon as the black hole relaxes back to the extremal black hole \cite{Hollowood:2020cou}. We will not be concerned with this very long time regime in this paper.

In order to calculate the entropy of the island saddle with the solution \eqref{kis}, the relevant terms in \eqref{box}, that contribute at leading order, are
\EQ{
&\frac{\pi c}{6\beta k}\cdot\frac{1-w_{\hat2}^+w_{\hat2}^-}{1+w_{\hat2}^+w_{\hat2}^-}\left(1-\frac{\beta k}{4\pi}\log w_{\hat2}^+\right)+\frac c{12}\log w_{\hat 2}^+
\,\longrightarrow \,S_\text{BH}\big(1+{\cal S}_{b_\alpha}\big)\ ,\\[5pt]
&-\frac c6\sum_j(-1)^j\log(w_{\hat 2}^-+\hat f_j^{-1})\,\longrightarrow \,4S_\text{BH}\sum_j(-1)^j{\cal S}_{b_{\text{max}(\alpha,j)}}\ .
}
The entropy only depends on $\log w_{\hat 2}^+$  at the leading order in $k$ wherein 
$\log w_{\hat 2}^+ \simeq \log \pmb{\omega}_{\hat 2}^+ ={\mathscr O}(k^{-1})$.
Hence, the entropy of the $\alpha$-labelled island saddle, $t>b_\alpha$, is
\EQ{
S_{\text{island}(\alpha)}=S_\text{no island}+S_\text{BH}\left\{1+{\cal S}_{b_\alpha}+4\sum_{j=1}^{2N}(-1)^j{\cal S}_{b_{\text{max}(j,\alpha)}}\right\}\ .
}

If the extremal entropy is not negligible, we must add $2S_0$ to the island saddle entropies, but for the present purpose we will mostly ignore it. Note that the islands do not lead to additional UV divergences because these are absorbed into a renormalization of $\phi_0$. The island saddles with $\alpha$ even, never have the lowest entropy and can therefore be ignored. This follows from the inequalities,
\EQ{
S_{\text{island}(2\alpha)}-S_{\text{island}(2\alpha\pm1)}=(-1\pm2)S_\text{BH}e^{-kt/2}(e^{kb_{2\alpha\pm1}/2}-e^{kb_{2\alpha}/2})>0\ .
}

\subsection{Both QES behind the shockwave}

Now let us consider the possibility that both QES are behind the shockwave. The generalized entropy is 
\EQ{
S_\text{gen.}(w^\pm_{\hat a})&=\sum_{\hat a}\Big(\frac{\phi(p_{\hat a})}{4G_N}-\frac c6\log\Omega_{\hat a}-\sum_j(-1)^{\hat a-j}\log \sigma_{\hat aj}\Big)\\ &-\sum_{\hat a<\hat b}(-1)^{\hat a-\hat b}\log \sigma_{\hat a\hat b}+S_\text{no island}\ ,\label{2qes}
}
where, for each QES, the dilaton and conformal factor are given in \eqref{box}. In the above equation, the spacetime intervals are
\EQ{
\sigma_{\hat a\hat b}=-(w^-_{\hat a}-w^-_{\hat b})(y^+_{\hat a}-y^+_{\hat b})\ ,\quad
\sigma_{\hat aj}=-(w_{\hat a}^-+\hat f_j^{-1})(y_{\hat a}^+-t-b_j)\ .
}
We are assuming that the order of the points on a Cauchy surface, from left to right, is $(p_{\hat1},p_{\hat2},p_1,p_2,\ldots,p_{2N})$. 

The position of the second QES $p_{\hat2}$ is again determined by equation \eqref{quf3}, and now the first QES $p_{\hat1}$ satisfies the same equation with the replacement $p_{\hat2}\leftrightarrow p_{\hat1}$,
\EQ{
&\frac{2\pi}{\beta k}\pmb{\omega}_{\hat 1}^+-\frac1{w_{\hat1}^--w_{\hat2}^-} -\sum_{j=1}^{2N}\frac{(-1)^j}{w_{\hat 1}^-+\hat f_j^{-1}}=0\ ,\\
&\frac{2\pi}{\beta k}w_{\hat1}^--\frac1{4\, \pmb{\omega}_{\hat1}^+}=0\ .
\label{quf4}
}
Again we have self-consistently discarded terms that are higher order in $k$.  Each QES has $2N$ possible saddle points labelled as $\alpha(\hat a)$ where $\hat a =1,2$. The solutions generalize \eqref{kis} with
\EQ{
w_{\hat a}^+=\frac{\beta kc_{\hat a}}{2\pi}\hat f_{\alpha({\hat a})} \,e^{\frac12 ky_{\alpha(\hat a)}^-}\ ,\qquad w_{\hat a}^-=\frac1{4c_{\hat a}\hat f_{\alpha(\hat a)}}\ .
\label{xit}
}
Here $c_{\hat a}=\frac34$ for $\alpha(\hat1)$ even or $\alpha(\hat2)$ odd and $c_{\hat a}=\frac1{12}$ for $\alpha(\hat1)$ odd or $\alpha(\hat2)$ even. Since $w_{\hat1}^+<w_{\hat2}^+$, we must have\footnote{${\pmb{\omega}_{\hat a}^+}$  and $w_{\hat a}^+$ only differ by order one terms in the exponent, and therefore on the time scales of interest, $\log \pmb{\omega}_{\hat1}^+ <
\log \pmb{\omega}_{\hat2}^+ $ $\implies$ $\log w_{\hat1}^+ <
\log w_{\hat2}^+ $.}
 $\alpha(\hat1)>\alpha(\hat2)$. 

We note that there is a term involving $(y^+_{\hat1}-y^+_{\hat2})$ in the second equation in \eqref{quf3} and in \eqref{2qes} which contributes in the special case that $\alpha(\hat1)=\alpha(\hat2)$. But this case yields an entropy that is always greater than the no-island entropy:
\EQ{
S_{\text{island}(\alpha,\alpha)}&=S_\text{no island}+2S_\text{BH}\big(1-{\cal S}_{b_\alpha}\big)>S_\text{no island}
}
and so is never dominant. Consequently from now on, we will assume that $\alpha({\hat2})\neq\alpha(\hat1)$.

Since we are taking both QESs to be behind the shockwave, the saddle points above only appear when $t>b_{\alpha(\hat1)}$ and they have entropy given by evaluating \eqref{box} in the small $k$ regime, yielding,
\EQ{
&S_{\text{island}(\alpha(\hat1),\alpha(\hat2))}=S_\text{no island}\\[5pt]&+S_\text{BH}\Big\{2+\sum_{\hat a=1}^2\Big({\cal S}_{b_{\alpha(\hat a)}}+4\sum_{j=1}^{2N}(-1)^{j-\hat a}{\cal S}_{b_{\text{max}(j,\alpha(\hat a))}}\Big)-4{\cal S}_{b_{\alpha(\hat1)}}\Big\}\ .
}
There is a zoo of possible saddles that can compete for the minimal entropy. However, 
many of the saddles can never have minimal entropy, given that $1\geq{\cal S}_{b_i}\geq{\cal S}_{b_j}$, for $i<j$. The following is a list of saddles for one and two intervals which can have minimal entropy. We write them in shorthand as a vector $(c_0,c_1,\ldots,c_{2N})$, 
\EQ{
S=S_\text{BH}\left(c_0+\sum_{j=1}^{2N}c_j{\cal S}_j\right)\ .
}
Using this notation we explicitly list  $S_\text{no island}$, the island entropy with one QES behind the shockwave $S_{\rm island (\alpha)}$, and  the entropy $S_{\rm island (\alpha(\hat 1), \alpha (\hat 2))}$ with both QESs behind the shockwave:
\EQ{
S_\text{no island}=(0,2,-2)\ ,\quad 
S_\text{island(1)}=(1,-1,2)\ ,\quad 
S_\text{island(21)}=(2,-1,-1)\ .
}
Then, for two intervals we have,
\EQ{
S_\text{no island}&=(0,2,-2,2,-2)\ ,\quad~~~\,
S_\text{island(1)}=(1,-1,2,-2,2)\ ,\\
S_\text{island(3)}&=(1,2,-2,-1,2)\ ,\quad~~\,
S_\text{island(21)}=(2,-1,-1,2,-2)\ ,\\
S_\text{island(41)}&=(2,-1,2,-2,-1)\ ,\quad
S_\text{island(43)}=(2,2,-2,-1,-1)\ .\label{twointerval}
}
Note that in the expressions above, we have left the cut off term implicit.

\subsection{Position of the QES and scrambling time}

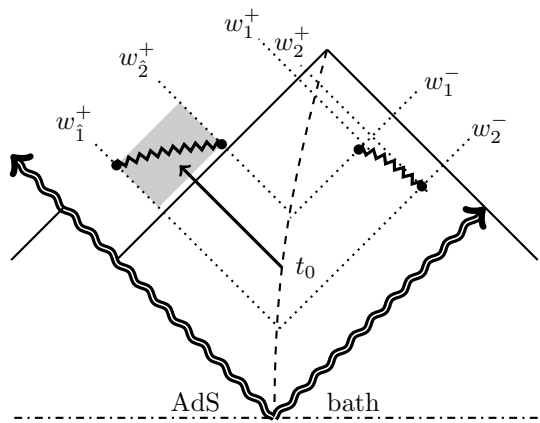
\begin{figure}[ht]
\begin{center}
\begin{tikzpicture} [scale=0.7]
\filldraw[color=black!20,draw=none] (-3,4.8) -- (-2.2,4) -- (-1,5.2) -- (-1.8,6) -- (-5,2.8); 
\draw[thick,dotted] (-3.4,5.2) -- (0.1,1.7);
\draw[thick,dotted] (-2.2,6.4) -- (0.3,3.9);
\draw[thick,dotted] (3.7,5.3) -- (0.1,1.7);
\draw[thick,dotted] (2.7,6.2) -- (0.3,3.8);
\node at (3.2,6.4) {\footnotesize $w^-_1$};
\node at (4.1,5.5) {\footnotesize $w^-_2$};
\draw[very thick,->] (0.15,2.85) -- (-1.8,4.8);
\node at (0.6,2.85) {\footnotesize $t_0$};
\node at (-3.8,5.6) {\footnotesize $w^+_{\hat1}$};
\node at (-2.6,6.8) {\footnotesize $w^+_{\hat2}$};
\draw[thick,dashed] (0,0) to[out=90,in=-105] (1,7);
\draw[thick] (1,7) -- (5,3);
\draw[thick,dash dot] (5,0)  -- (-5,0);
\draw[double,decorate,very thick,->,decoration={snake,amplitude=0.03cm}] (0,0) -- (-5,5);
\draw[double,decorate,very thick,->,decoration={snake,amplitude=0.03cm}] (0,0) -- (4,4);
\draw[decorate,very thick,decoration={zigzag,segment length=1.5mm,amplitude=0.5mm}] (-3,4.8) to[out=20,in=-175] (-1,5.2);
\filldraw[black] (-1,5.2) circle (2.5pt);
\filldraw[black] (-3,4.8) circle (2.5pt);
\draw[decorate,very thick,decoration={zigzag,segment length=1.5mm,amplitude=0.5mm}] (1.6,5.1) to[out=-10,in=145] (2.8,4.4);
\filldraw[black] (1.6,5.1) circle (2.5pt);
\filldraw[black] (2.8,4.4) circle (2.5pt);
\draw[thick,dotted] (1.8,5.1) -- (-0.3,7.2);
\node at (-0.6,7.5) {\footnotesize $w^+_1$};
\draw[thick,dotted] (2.9,4.3) -- (0.5,6.7);
\node at (0.35,7.15) {\footnotesize $w^+_2$};
\draw[thick] (-5,3) -- (-4,4);
\draw[thick] (-3,3) -- (1,7);
\node at (1.5,0.3) {\footnotesize bath};
\node at (-1.5,0.3) {\footnotesize AdS};
\end{tikzpicture}
\caption{\footnotesize The $w^+$ coordinates of the QESs determine when a null ray from the boundary is in the island of an interval in the bath. The shaded area is the island-$(21)$ of a single interval in the bath at time $t$. Note, also, the relation between the $w^-$ coordinates branch points in the bath and the $w^+$ coordinates of the QESs.}
\label{fig7} 
\end{center}
\end{figure}

It is important to know when an in-going null ray beginning on the boundary at time $t_0$ lies in the island of a saddle at time $t$. Given the coordinates of the QESs in \eqref{xit}, the condition is
\EQ{
\frac{\beta kc_{\alpha(\hat 1)}}{2\pi}e^{\tfrac12 ky_{\alpha(\hat 1)}^-}\,\hat f_{\alpha(\hat1)}\,<\hat f(t_0)<\frac{\beta kc_{\alpha(\hat 2)}}{2\pi}e^{ \tfrac12 ky_{\alpha(\hat 2)}^-}\,\hat f_{\alpha(\hat2)} \ ,
} 
where the lower bound is only important if both QES are behind the shockwave. The bounds here, using our approximations, are
\EQ{
t_0=t-b_{\alpha(\hat a)}-e^{k(t-b_{\alpha(\hat a)})/2}\frac{\beta }{2\pi}\log\left(\frac{2\pi }{\beta  k c_{\alpha(\hat a)}}e^{-k(t-b_{\alpha(\hat a)})/2}\right)\ .
}
The final term here is an expression of the {\it scrambling time\/}, the time it takes to recover information dropped into an old black hole as we shall verify later. This term is sub-leading in the small $k$ limit, where $t_0$ and the $b_j$ are order $k^{-1}$. Interestingly, if we ignore the small correction from the scrambling time and partner the in-going ray at $t_0=t-b_{\alpha(\hat a)}$ on the boundary with an out-going ray, then the latter hits the branch point $p_{\alpha(\hat a)}$ in the bath at time $t$ just as the in-going one hits the QES $p_{\hat a}$, as shown in figure \ref{fig7}.

\section{Page curve and early/late correlations}\label{s4}

The Page curve is simply the entropy of a single interval $[0,t]$ in the bath at time $t$  that collects all the Hawking radiation emitted during the life of the evaporating black hole up to time $t$, i.e.~in the temporal interval $\langle 0,t\rangle$.\footnote{We use angle brackets to distinguish a temporal interval from a spatial one. Importantly, when we write $[0,t]$ we mean the limit $\epsilon\to0$ of $[\epsilon,t]$, to emphasize that in the holographic interpretation the boundary system is {\it not\/} included.} Note that the point $p_2$ lies on the wave front of the Hawking radiation. The curve involves a competition between the no-island and island-$(1)$ saddles:
\EQ{
S_\text{Page cuve}(t)&=S_\text{BH}\min\big(2{\cal S}(t),1-{\cal S}(t)\big)\\[5pt]
&=S_\text{BH}\min\big(2-2e^{-kt/2},e^{-kt/2}\big)\ .
}
The Page time occurs when the no-island ceases to dominate and the island saddle takes over. This occurs at
\EQ{
t_\text{Page}=\frac2k\log\frac32\ .
}
If we do not neglect the extremal entropy and write $S_0=\xi S_\text{BH}$, then the Page time is increased to $t_\text{Page}=\frac 2k\log\frac3{2(1-\xi)}$. Clearly, if $\xi>1$ then our approximation regime breaks down and full Bessel function expression for $\hat f$ will be needed. In the following, for simplicity, we will suppose that $\xi$ is small and can be neglected.

\subsection{Mutual information}

\begin{figure}[ht]
\begin{center}
\begin{tikzpicture} [scale=0.7]
\filldraw[color=black!45,draw=none] (0,0) -- (4,4) -- (3.5,4.5) -- (-0.5,0.5) -- (0,0); 
\filldraw[color=black!25,draw=none] (-0.5,0.5) -- (3.5,4.5) -- (1.85,6.15) -- (-2.05,2.15) -- (-0.5,0.5); 
\draw[thick,dashed] (0,0) to[out=90,in=-105] (1,7);
\draw[thick] (1,7) -- (5,3);
\draw[thick,dash dot] (5,0)  -- (-5,0);
%
%
\draw[double,decorate,very thick,->,decoration={snake,amplitude=0.03cm}] (0,0) -- (-5,5);
\draw[double,decorate,very thick,->,decoration={snake,amplitude=0.03cm}] (0,0) -- (4,4);
\draw[thick] (-5,3) -- (-4,4);
\draw[thick] (-3,3) -- (1,7);
\draw[decorate,very thick,decoration={zigzag,segment length=1.5mm,amplitude=0.5mm}] (0.5,4.75) to[out=-5,in=140] (3.7,3.7);
\filldraw[black] (0.5,4.75) circle (2.5pt);
\filldraw[black] (3.15,4.15) circle (2.5pt);
\filldraw[black] (3.7,3.7) circle (2.5pt);
%
%
\node at (1.5,0.3) {\footnotesize bath};
\node at (-1.5,0.3) {\footnotesize AdS};
\node at (1.3,4) {\footnotesize $A$};
\node at (2.2,2.7) {\footnotesize $B$};
%
\node at (0.2,4.75) {\footnotesize $t$};
\node at (3,1.1) (a1) {\footnotesize $t_\text{Page}$};
\draw[thick,dotted] (0.1,1.1) -- (a1);
\end{tikzpicture}
\caption{\footnotesize A scenario to measure the mutual information of the early and late Hawking radiation emitted in time intervals $B=\langle 0,t_\text{Page}\rangle$ and $A=\langle t_\text{Page},t\rangle$ by collecting the radiation in appropriate spatial intervals on a Cauchy surface.}
\label{fig3} 
\end{center}
\end{figure}
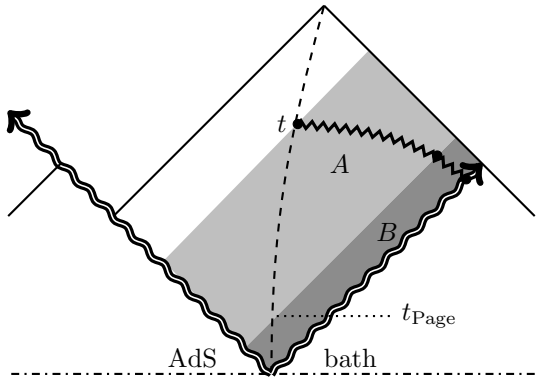

We can compute the mutual information between the early and late radiation emitted in time intervals  $\langle 0,t_\text{Page}\rangle$ and $\langle t_\text{Page},t\rangle$, for $t>t_\text{Page}$. In order to do this, one collects the two subsets in spatial intervals $B=[t-t_\text{Page},t]$ and $A=[0,t-t_\text{Page}]$, respectively, at time $t$: see figure \ref{fig3}.
In order for the result to be UV safe, we suppose the two intervals are slightly separated by more than the UV cut off. The mutual information is 
\EQ{
I_{A,B}=S_A+S_B-S_{AB}\ .
}
It is easy to see that $B$ is at the island-$(1)$/no-island threshold, while $AB$ is always in its island-$(1)$ saddle. The interval $A$ has a transition from its no-island to island saddle at a time $t=\frac2k\log\frac92$:
\EQ{
S_A&=S_\text{BH}\min\Big(\frac43-2e^{-kt/2},\frac23+e^{-kt/2}\Big)\ ,\\[5pt]
S_B&=\frac23S_\text{BH}\ ,\qquad 
S_{AB}=S_\text{BH}e^{-kt/2}\ .
\label{gur}
}
Hence, the mutual information increases as $t$ increases from $t_\text{Page}$, but then plateaus for $t>\frac2k\log\frac92$:
\EQ{
I_{A,B}=S_\text{BH}\min\Big(2-3e^{-kt/2},\frac43\Big)\ .
\label{bee}
}

\subsection{Quantum correlations}

The fact that the early and late modes have non-vanishing mutual information means that they are correlated but  this does not discriminate between classical and quantum correlations. 
One way to measure genuine quantum correlations, is via the behaviour of the conditional entropies \cite{Cerf:1995sa}
\EQ{
S_{A|B}=S_{AB}-S_B\ ,\quad S_{B|A}=S_{AB}-S_A\ .
}
These must be non-negative in a classical system and so their negativity is a measure of quantum correlations. In the present case,
\EQ{
S_{A|B}=S_\text{BH}\Big(e^{-kt/2}-\frac23\Big)\ ,\qquad
S_{B|A}=-S_\text{BH}\min\Big(\frac23,\frac43-3e^{-kt/2}\Big)\ .
}
Recalling that $t>t_\text{Page}$, the former is always negative and the latter becomes negative for $t>\frac2k\log\frac92$.

We can consider the same correlation measures on other subsets of the radiation. For example in figure \ref{fig10}, we show the conditional entropies for two subsets of the radiation $\langle0,t\rangle$ and $\langle t,4t_\text{Page}\rangle$ with varying $t$. The existence of genuine quantum correlation is clear.

\pgfdeclareimage[interpolate=true,width=7cm]{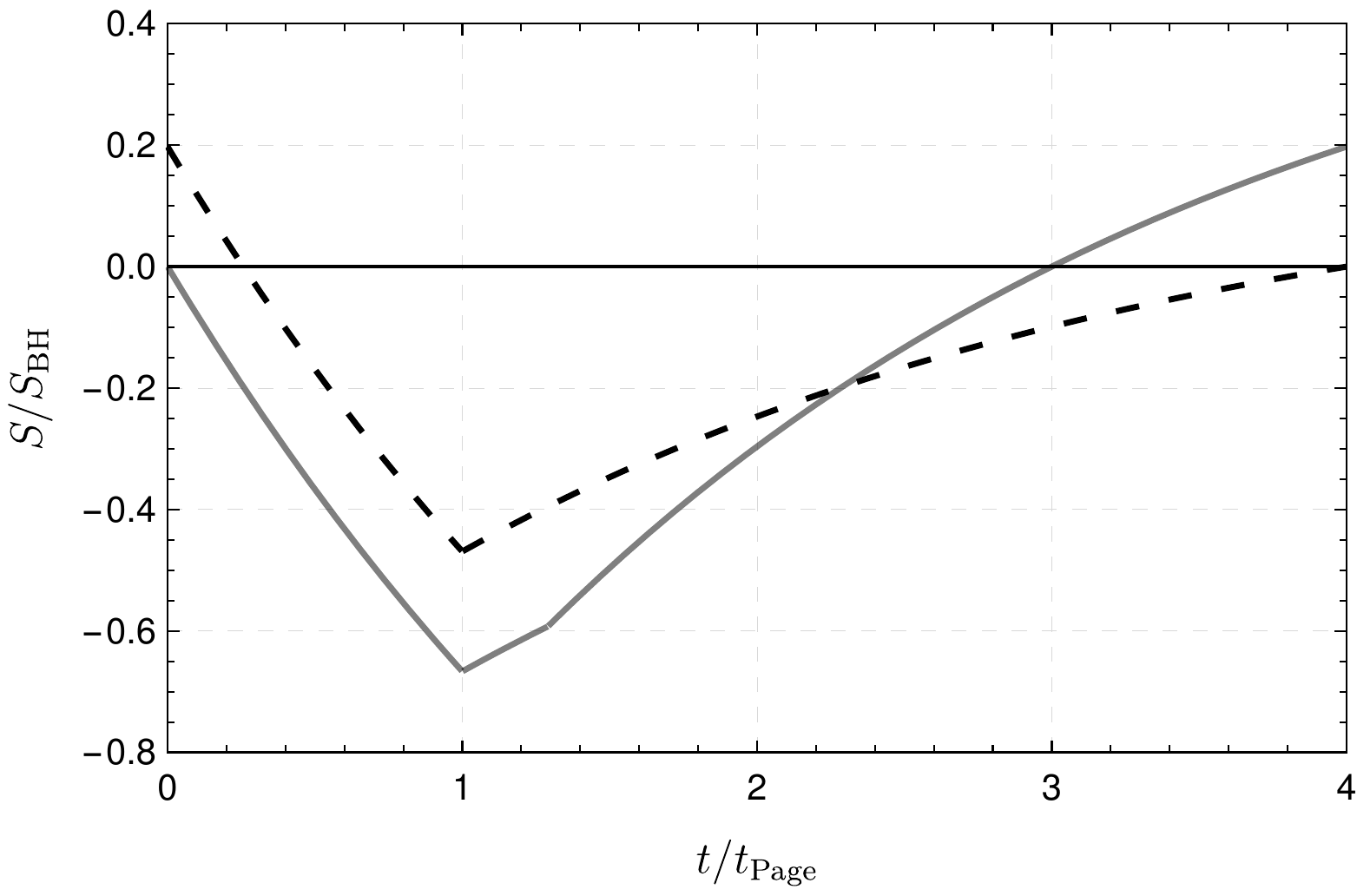}{NCE}
\pgfdeclareimage[interpolate=true,width=7cm]{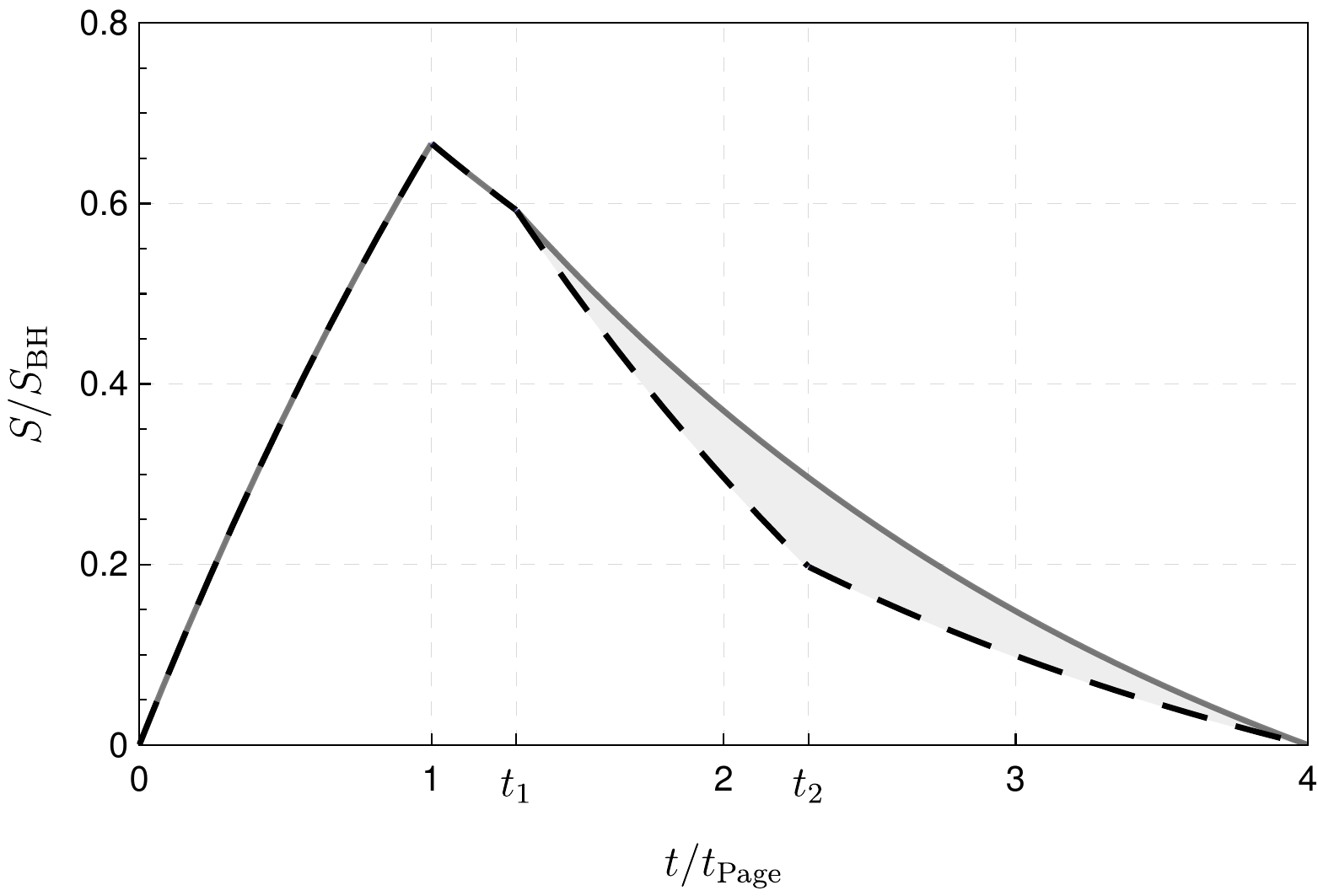}{Esq4}
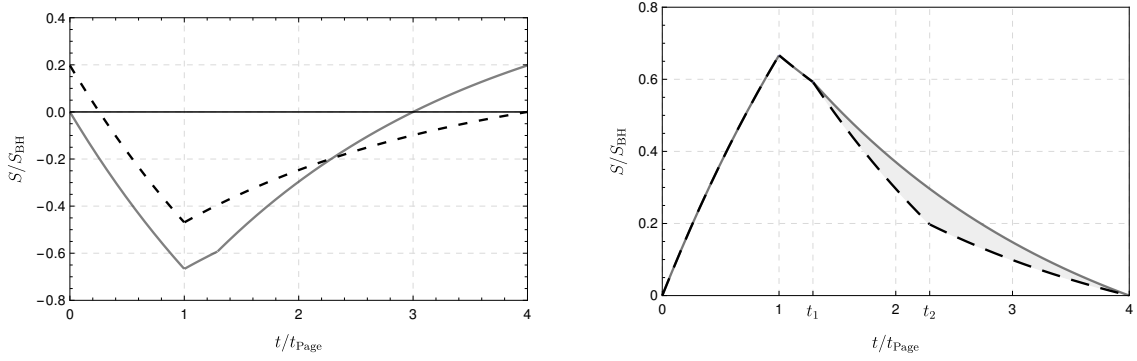
\begin{figure}
\begin{center}
\begin{tikzpicture}[scale=1]
\pgftext[at=\pgfpoint{0cm}{0cm},left,base]{\pgfuseimage{NCE}}
\pgftext[at=\pgfpoint{8cm}{0cm},left,base]{\pgfuseimage{Esq4}}
\end{tikzpicture}
\caption{\small {\it Left:\/} A plot of $S_{A|B}$ (dashed) and $S_{B|A}$ (solid) for the two temporal subsets $B = \langle 0, t \rangle$ and $A = \langle t , 4t_\text{Page}\rangle$. Notice that at least one of them is negative at each $t$ indicating the presence of quantum correlations. {\it Right:\/} for the same regions a plot of the upper and lower bounds on the squashed entanglement (assuming the UV cut off term is sufficiently small). Thus proves the existence of entanglement between $A$ and $B$.}
\label{fig10}
\end{center}
\end{figure}

Measuring entanglement between two subsystems $A$ and $B$ in quantum mechanics is not simple or obvious when the state on $AB$ is not pure. An excellent measure of quantum correlations is provided by the  {\it squashed entanglement\/} \cite{Tucci,CW,CL} (see also \cite{Hayden:2011ag,Takayanagi:2017knl})  defined as 
\EQ{
E^\text{sq.}_{A,B}=\frac12\min_C I_{A,B|C}\ ,
\label{her}
}
involving the {\it conditional mutual information\/}
\EQ{
I_{A,B|C}=I_{A,BC}-I_{A,C}\ .
}
Here, $C$ is {\it any\/} additional quantum system appended to $A$ and $B$. Unfortunately, this latter feature renders it impractical to calculate the squashed entanglement. However, it can be bounded both from above and below \cite{CL} by quantities that are calculable:
\EQ{
\max(0,-S_{A|B},-S_{B|A})\leq E^\text{sq.}_{A,B}\leq\frac12I_{A,B}\ .
\label{pap}
}
Notice that the lower bound involves minus the conditional entropies, so a non-trivial bound occurs when one or both of the conditional entropies are negative, indicating quantum entanglement. In the QFT setting, the lower bound can be made UV safe, if $A$ and $B$ (but not $C$) are taken to be adjacent. 

Let us consider two subsets of radiation $\langle 0,t\rangle$ and $\langle t,T\rangle$ collected in bath regions $A=[0,T-t]$ and $B=[T-t,T]$ at time $T$. One can compute the upper and lower bounds of the squashed entanglement, $E^\text{sq.}_\gtrless$, as in \eqref{pap}. There are four temporal regions:
\EQ{
&t\leq t_\text{Page} \qquad\, \qquad E^\text{sq.}_<=E^\text{sq.}_>=2{\cal S}(t)\ ,\\[5pt]
&t_\text{Page}<t\leq t_1 \qquad E^\text{sq.}_<(-S_{B|A})=E^\text{sq.}_>=1-{\cal S}(t)\ ,\\[5pt]
&t_1<t\leq t_2 \qquad~~~\, \begin{cases}E^\text{sq.}_<(-S_{B|A})=-1+3{\cal S}(T)-2{\cal S}(t)\ , &\\ E^\text{sq.}_>=\tfrac32({\cal S}(T)-{\cal S}(t))\ , &\end{cases}\\[5pt]
&t>t_2 \qquad\qquad~~~~ \begin{cases}E^\text{sq.}_<(-S_{A|B})={\cal S}(T)-{\cal S}(t)\ ,& \\ E^\text{sq.}_>=\tfrac32({\cal S}(T)-{\cal S}(t))\ ,&\end{cases}
}
where $t_1=T-\frac{\log3}{\log3/2}t_\text{Page}$ and $t_2=T-\frac{\log2}{\log3/2}t_\text{Page}$. Figure \ref{fig10} shows the upper and lower bounds for our example with $T=4t_\text{Page}$.

Even though we cannot calculate the squashed entanglement directly, we can gain some intuition by choosing $C$ in \eqref{her} to be another subset of the Hawking radiation. In figure \ref{fig11} we illustrate that, when the interval $C$ is distinct, i.e.~has no overlap with $A$ and $C$, that
\EQ{
 I_{A,B|C}\geq I_{A,B}\ .
}
So, even though we cannot claim that $E^{\text{sq.}}_{A,B}=\frac12 I_{A,B}$, this is an  indication that the mutual information we have calculated is a good measure of entanglement since we managed to ``squash away'' some of the non-quantum correlations by choosing a suitable $C$ of the Hawking radiation. Moreover, since the mutual information is an upper bound for the squashed entanglement, this is another indication that the correlations amongst subsets of the Hawking radiation are due to entanglement and are not just classical.

\pgfdeclareimage[interpolate=true,width=7cm]{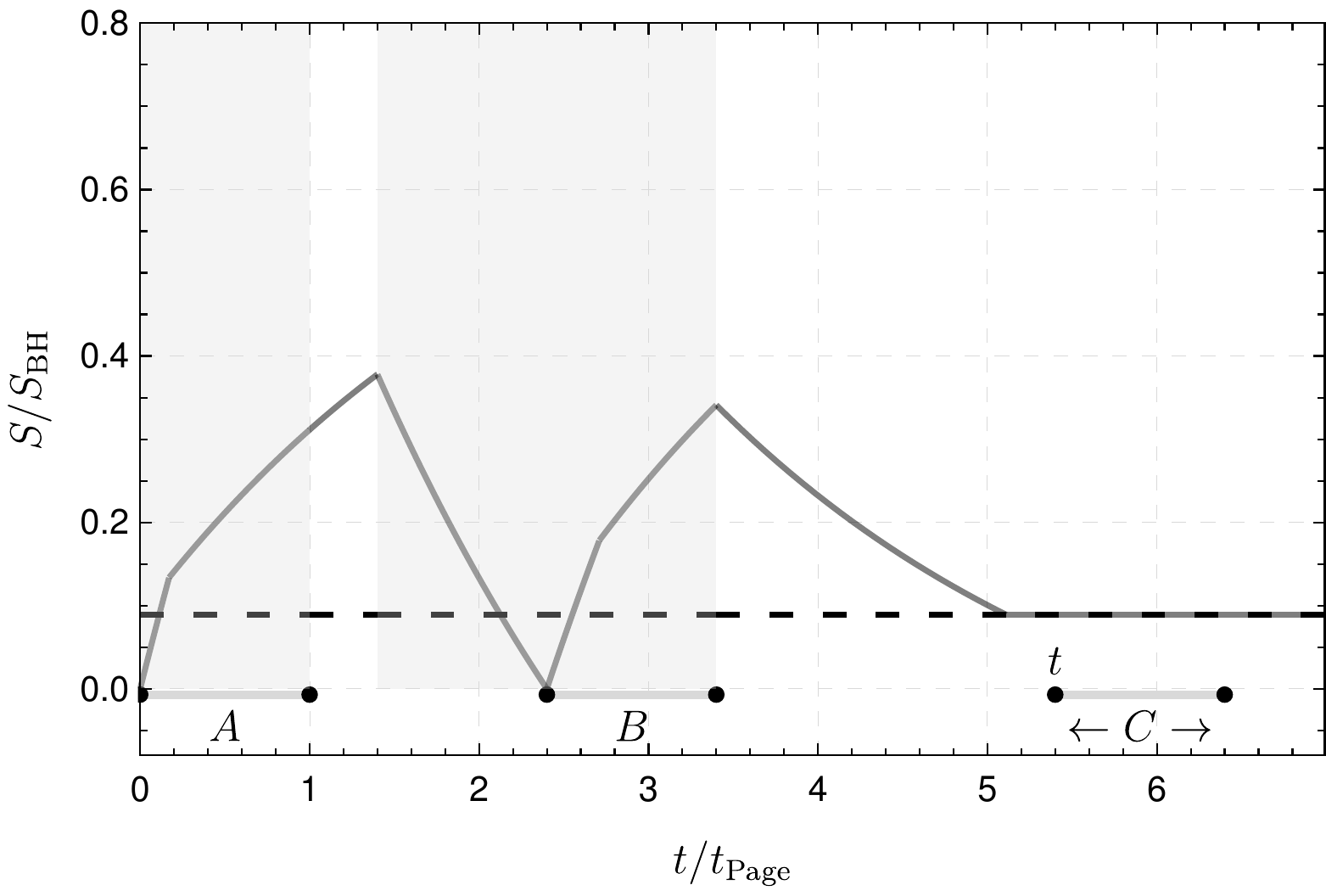}{HR1}
\pgfdeclareimage[interpolate=true,width=7cm]{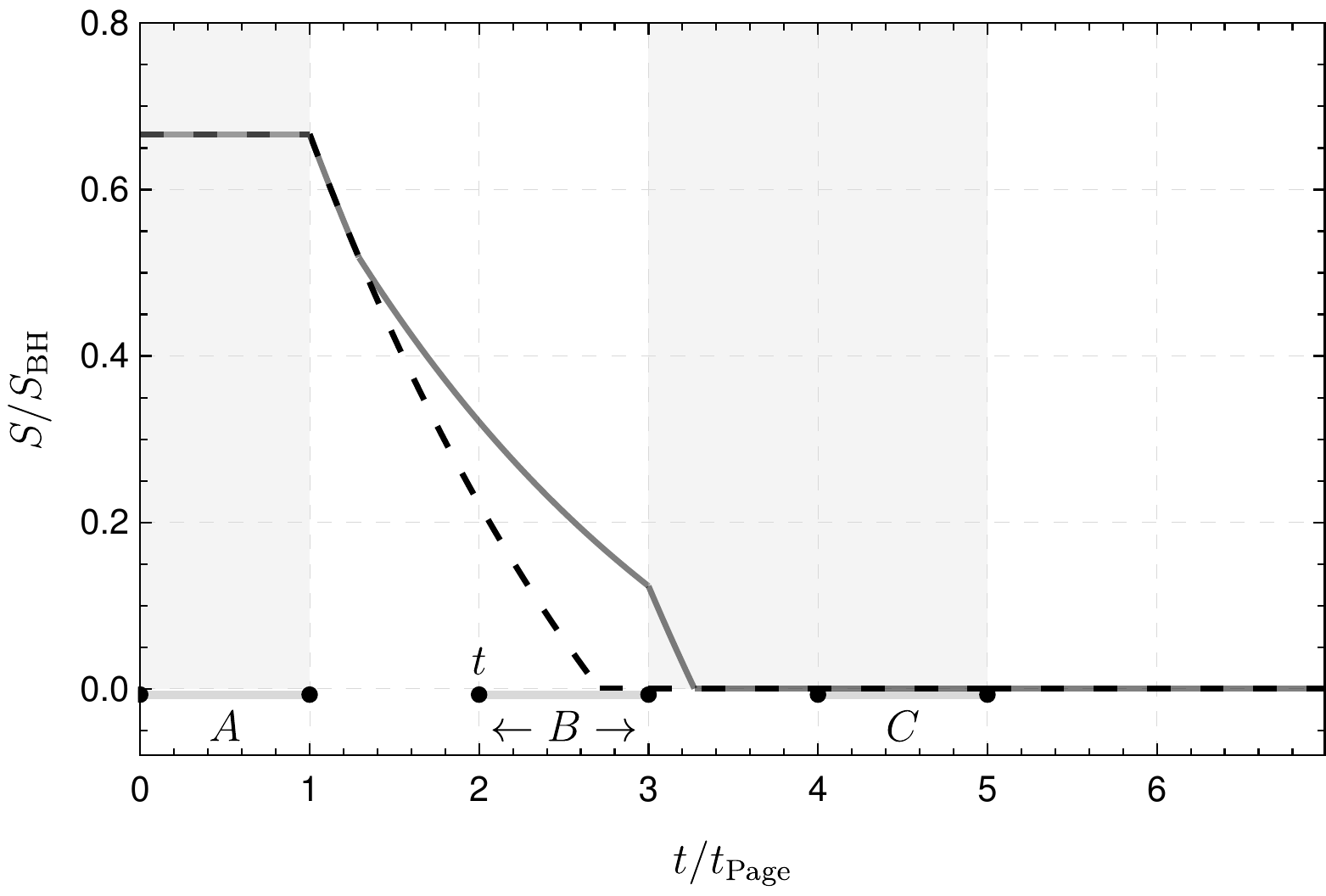}{HR2}
\begin{figure}
\begin{center}
\begin{tikzpicture}[scale=1]
\pgftext[at=\pgfpoint{0cm}{0cm},left,base]{\pgfuseimage{HR1}}
\pgftext[at=\pgfpoint{8cm}{0cm},left,base]{\pgfuseimage{HR2}}
\end{tikzpicture}
\caption{\small These plots show the mutual information $I_{A,B}$ (dotted) and conditional mutual information $I_{A,B|C}$ (continuous) for three intervals $A,B,C$ in the Hawking radiation collected at a fixed time $8 t_{\text{Page}}$. {\it Left:}~with $A$ and $B$ fixed with $C$ moving and {\it Right:}~with $A$ and $C$ fixed with $B$ moving. The shaded regions indicates where the intervals overlap. Notice that we get $I_{A,B|C}=0$ when $C=A$ or $B$, as expected from its definition. The non-negativity of $I_{A,B}$ and $I_{A,B|C}$ are checks of sub-additivity and strong sub-additivity, respectively.}
\label{fig11}
\end{center}
\end{figure}
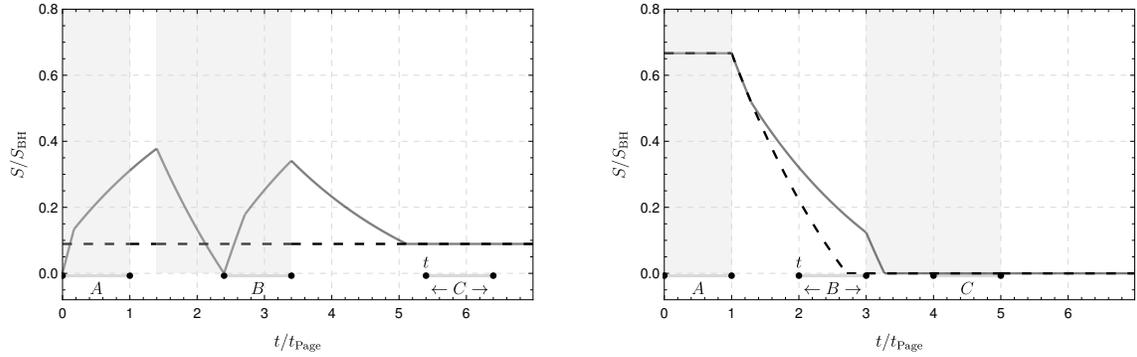

\subsection{Entropy consistency conditions}

The von Neumann entropies of entanglement must satisfy various consistency conditions in a generic quantum system, including sub-additivity and the Araki-Lieb inequality. Firstly, if we assume that the UV cut-off is sub-leading compared with the order $k^{-1}$ contributions to the entropy, then one can verify that the Araki-Lieb triangle inequality,
\EQ{
S_{AB}\geq \big|S_A-S_B\big|\ ,
}
is satisfied by our subset of modes $\langle 0,t_\text{Page}\rangle$ and $\langle t_\text{Page},t\rangle$. We can also make the same check for other subsets; for example in figure \ref{fig12}, we show $S_{AB}$, $S_A+S_B$ and $|S_A-S_B|$ for subsets $\langle 0,t-\frac12t_\text{Page}\rangle$ and $\langle t,5 t_\text{Page}\rangle$ as a function of $t$.

\pgfdeclareimage[interpolate=true,width=7cm]{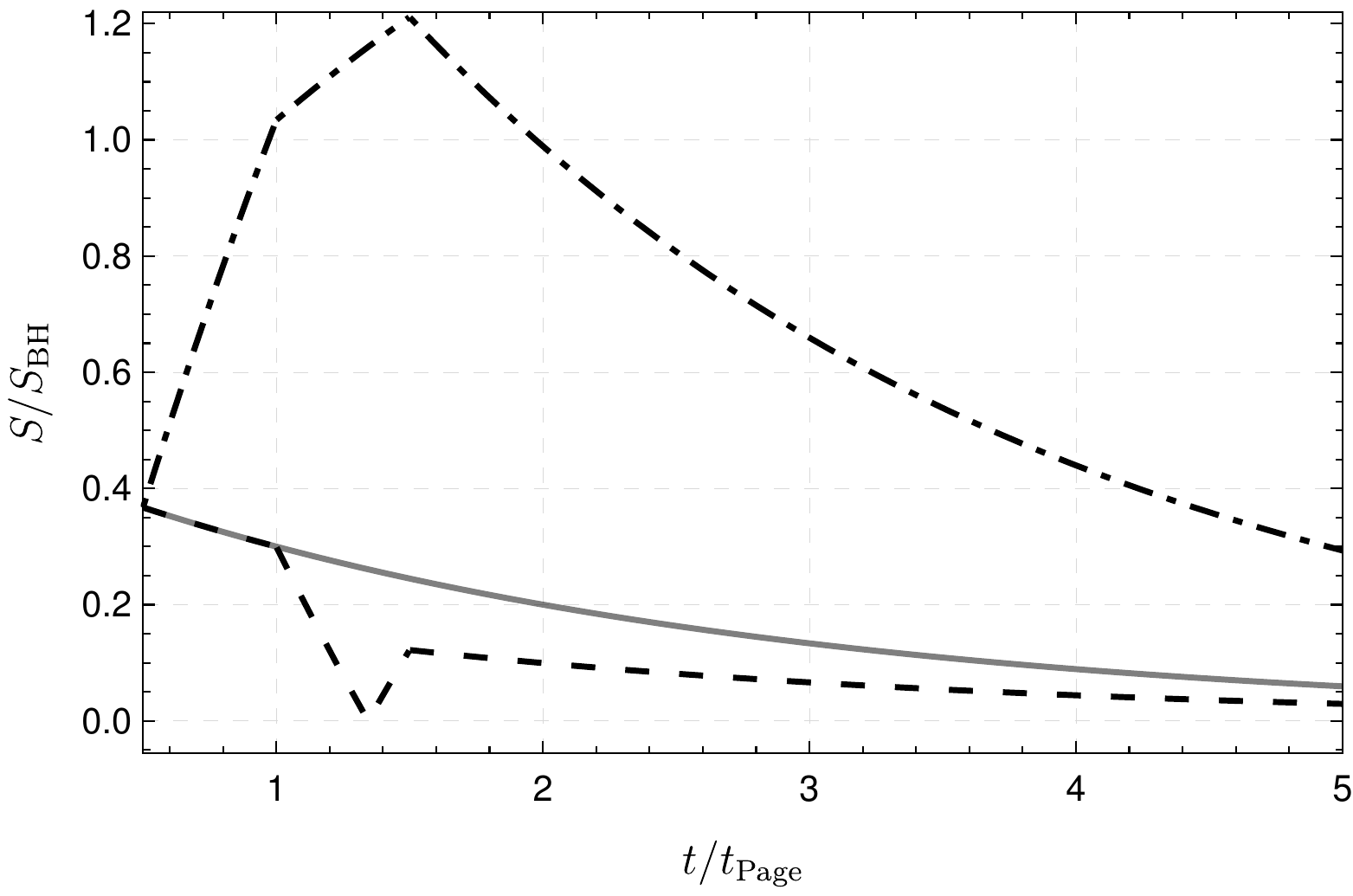}{AR}
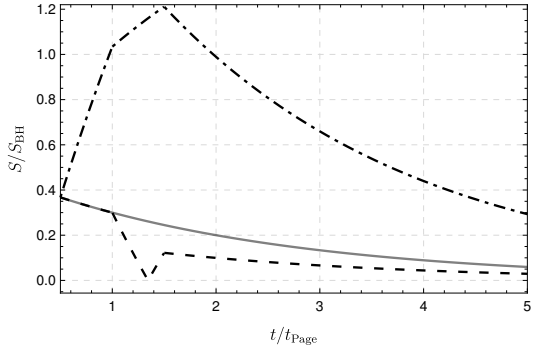
\begin{figure}
\begin{center}
\begin{tikzpicture}[scale=1]
\pgftext[at=\pgfpoint{0cm}{0cm},left,base]{\pgfuseimage{AR}}
\end{tikzpicture}
\caption{\small Check of the Araki-Lieb triangle inequality and sub-additivity of entanglement entropy, $S_A+S_B\geq S_{AB}\geq|S_A-S_B|$, with $S_{AB}$ (continuous), $|S_A-S_B|$ (dashed) and $S_A+S_B$ (dot-dashed). Here we considering, for example, the following two subsets $B = \langle 0, t-\tfrac12 t_{\text{Page}} \rangle$ and $A = \langle t , 5t_\text{Page}\rangle$.}
\label{fig12}
\end{center}
\end{figure}

The next consistency condition is {\it sub-addivity\/}; namely the non-negativity of the mutual information $I_{A,B}\geq0$. This is clearly satisfied for our subsets $\langle 0,t_\text{Page}\rangle$ and $\langle t_\text{Page},t\rangle$ in \eqref{bee}. It is also satisfied for the choices of subsets in figures \ref{fig11} and \ref{fig12}.

The final consistency condition we will consider is {\it strong sub-additivity\/} which can be stated in terms of non-negativity of the conditional mutual information, or
\EQ{
I_{A,BC}\geq I_{A,C}\ .
}
This is clearly satisfied in our choice of subsets  $\langle 0,t_\text{Page}\rangle$ and $\langle t_\text{Page},t\rangle$ because as $t$ increases, the subset $B$ is becomes larger, in other words we can think of this as adding a new subset of the radiation $B\to BC$, and the mutual information is decreasing. One can check that strong sub-additivity is also satisfied in the example of figure \ref{fig11} and for other choices of subsets.

To summarize, we do not have general proofs that all the necessary consistency conditions are satisfied by the entropies, but in all checks have tried they are seen to be satisfied.

\section{Hunt the purifier}\label{s5}

The implication of Page's argument \cite{Page:1993wv} for the behaviour of the entropy of the Hawking radiation is that a late mode $B$ of the radiation must be entangled with an early mode $R_B$, in order that the initial pure state evolves ultimately back to a pure state when the black hole evaporates away. In this section, we will calculate the mutual information of a small subset of late modes $B$ with subsets of early modes and by maximizing the latter, identify where $R_B$ lies within the early radiation. But then we shall turn the scenario around and ask which future modes $B$ is correlated with.

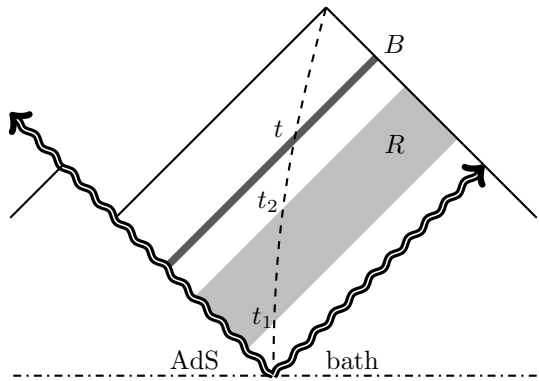
\begin{figure}[ht]
\begin{center}
\begin{tikzpicture} [scale=0.7]
\filldraw[color=black!25,draw=none] (-0.5,0.5) -- (3.5,4.5) -- (2.5,5.5) -- (-1.5,1.5) -- (-0.5,0.5); 
\filldraw[color=black!65,draw=none] (-2,2) -- (2,6) -- (1.9,6.1) -- (-2.1,2.1) -- (-2,2); 
\draw[thick,dashed] (0,0) to[out=90,in=-105] (1,7);
\draw[thick] (1,7) -- (5,3);
\draw[thick,dash dot] (5,0)  -- (-5,0);
\draw[double,decorate,very thick,->,decoration={snake,amplitude=0.03cm}] (0,0) -- (-5,5);
\draw[double,decorate,very thick,->,decoration={snake,amplitude=0.03cm}] (0,0) -- (4,4);
\draw[thick] (-5,3) -- (-4,4);
\draw[thick] (-3,3) -- (1,7);
\node at (1.5,0.3) {\footnotesize bath};
\node at (-1.5,0.3) {\footnotesize AdS};
\node at (2.3,6.3) {\footnotesize $B$};
\node at (2.3,4.4) {\footnotesize $R$};
\node at (0.1,4.7) {\footnotesize $t$};
\node at (-0.1,3.3) {\footnotesize $t_2$};
\node at (-0.2,1.1) {\footnotesize $t_1$};
\end{tikzpicture}
\caption{\footnotesize A small set of late Hawking modes $B$ and a subset of the early Hawking radiation $R$. We are interested in the correlation of $B$ and the smallest region $R$ with the maximum correlation.}
\label{fig4} 
\end{center}
\end{figure}

\subsection{Searching for $R_B$ in the past}

To this end, we pick a subset of Hawking modes $B$ emitted in a small temporal interval $\langle t,t+\delta\rangle$ captured in a spatial interval $[0,\delta]$ at time $t+\delta$. Note by small, we mean that the dimensional quantity $k\delta$ is small.
The question is, what is the smallest subset of the earlier radiation $R$ emitted between times $\langle t_1,t_2\rangle$, captured in spatial interval $[t+\delta-t_2,t+\delta-t_1]$ at time $t+\delta$, that has maximal mutual information at ${\mathscr O}(k^{-1})$ with the late modes $B$. This provides a way of identifying where the purifier $R_B$ of the modes $B$ are located in the earlier Hawking radiation

The first point to make is that $B$ is a narrow interval, so it is dominated by its no-island saddle, so
\EQ{
S_B=k\delta S_\text{BH}e^{-kt/2}\ .
}
In order for there to be non-vanishing mutual information, $R$ must be dominated by an island saddle. In particular, this requires the island-$(1)$ saddle to dominate and 
\EQ{
2e^{-kt_1/2}-\tfrac32e^{-kt_2/2}>1\ .
\label{tut}
}
However, when $R$ has an island, the combined subsystem $BR$ can be dominated by the  island-$(3)$ saddle and the mutual information vanishes because
\EQ{
S_{B,\text{no island}}+S_{R,\text{island}(1)}=S_{BR,\text{island}(3)}\ .
}
A non-vanishing $I_{B,R}$ only arises when $BR$ switches over to its  island-$(1)$  saddle. The competition between these two  determines the mutual information evaluated at time $t+\delta$ (with $R$ at its island-$(1)$ saddle)
\EQ{
I_{B,R}=S_\text{BH}\max\big(0,-3e^{-k(t+\delta)/2}+4e^{-kt/2}-e^{-kt_2/2}\big)\,.
}
Since $\delta$ is small, for $I_{B,R}$ to be non-vanishing we require
\EQ{
t_2>t-3\delta
}
and in order that the mutual information is a maximum requires that the early interval is adjacent to the late interval, i.e.~$t_2=t$, giving
\EQ{
I_{B,R}=\frac32S_B\ ,
\label{frr}
}
assuming that the cut off term is negligible and $\delta$ is small.

Notice that \eqref{frr} is independent of $t_1$ and so to identify where the entangled partner modes $R_B$ are, we should take $t_1$ as large as possible compatible with the constraint \eqref{tut} with $t_2=t$. This determines
\EQ{
t_1=\frac2k\log\frac4{2+3e^{-kt/2}}\ .
}
So the modes that are correlated with the late modes localized in the small interval $B=\langle t,t+\delta\rangle$ lie de-localized in the large interval of earlier modes
\EQ{
R_B\subset\Big\langle\frac2k\log\frac4{2+3e^{-kt/2}},t\Big\rangle\ .
\label{puc}
}
As a consistency check, note that above requires that $t>t_\text{Page}$ which is what we would intuitively expect: the black hole must be old. In the next subsection will see that the maximum mutual information is $\frac32S_B$ and not $2S_B$ because $B$ is also entangled with later modes.

\subsection{Searching for $R_B$ in the future}

We can reverse the logic of the last section, and consider the same scenario but where now $R$ are later Hawking modes. So now we have $t_1>t$ and the radiation corresponding to $R$ and $B$ are collected in intervals $[0,t_2-t_1]$ and $[t_2-t-\delta,t_2-t]$, respectively, at time $t_2$. One finds that in order to maximize the mutual information $I_{R,B}$, we need to begin collecting radiation immediately so that $t_1=t+\delta$. 

Then the question is, how small can $t_2$ be whilst maintaining maximum mutual information $I_{R,B}$. If the modes $B$ are early, i.e.~$t<t_\text{Page}$, the maximum $I_{R,B}$ is achieved when $S_R$ switches from its no-island to island-$(1)$  saddle. This occurs when \eqref{tut} is satisfied with $t_1=t$, which means the minimum $t_2$ is
\EQ{
t_2=\frac2k\log\frac3{4e^{-kt/2}-2}\ .
}
Since $S_{RB}$ is also in its island-$(1)$ saddle, we have, for small $\delta$,
\EQ{
I_{R,B}=2S_B\ ,
}
where we are assuming the cut off is negligible.

When $t>t_\text{Page}$, the r\^oles of the island-$(1)$ saddles of $R$ and $RB$ change to the $(21)$ and $(41)$ island saddles, respectively. Now the minimum $t_2$ is determined by when these saddles dominate over the non-island saddle, i.e.~when
\EQ{
t_2=t+\frac2k\log 3\ .
}
In this case,
\EQ{
I_{B,R}=\frac12S_B\ .
}
This is a very satisfying result because we know that for $t>t_\text{Page}$, $B$ is correlated with the early radiation as in \eqref{frr}, i.e.~$I_{B,R}=\frac32S_B$. So, overall, $S_B$ is correlated with the rest of the Hawking radiation with a mutual information $2S_B$, which is what one expect if $B$ is entangled with the rest of the radiation and the overall state is pure, at least approximately so when the extremal entropy is negligible.

\section{Entanglement-monogamy problem}\label{s6}

One way to present the entanglement-monogamy problem is simply the fact that unitarity requires for an old black hole that the Hawking radiation has a falling entropy, in other words, its state is becoming purer, whilst at the same time it must be entangled with modes behind the horizon in order to have a smooth geometry at the horizon. This seems contradictory.

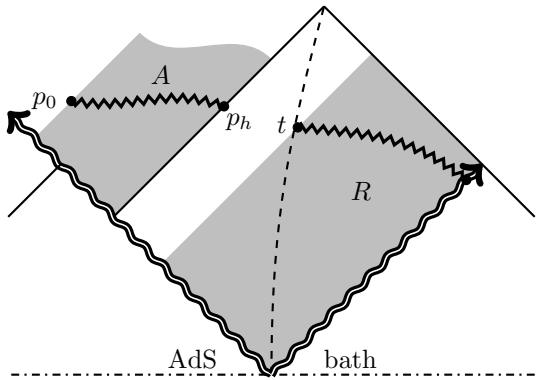
\begin{figure}[ht]
\begin{center}
\begin{tikzpicture} [scale=0.7]
\filldraw[color=black!25,draw=none] (0,0) -- (4,4) -- (1.9,6.1) -- (-2.1,2.1) -- (0,0); 
\filldraw[color=black!25,draw=none] (-3,3) -- (0,6) to[out=135,in=-45] (-2.5,6.5) -- (-4.5,4.5) -- (-3,3); 
\draw[thick,dashed] (0,0) to[out=90,in=-105] (1,7);
\draw[thick] (1,7) -- (5,3);
\draw[thick,dash dot] (5,0)  -- (-5,0);
%
%
\draw[double,decorate,very thick,->,decoration={snake,amplitude=0.03cm}] (0,0) -- (-5,5);
\draw[double,decorate,very thick,->,decoration={snake,amplitude=0.03cm}] (0,0) -- (4,4);
\draw[thick] (-5,3) -- (-4,4);
\draw[thick] (-3,3) -- (1,7);
\draw[decorate,very thick,decoration={zigzag,segment length=1.5mm,amplitude=0.5mm}] (0.5,4.7) to[out=-5,in=140] (3.7,3.7);
\filldraw[black] (0.5,4.7) circle (2.5pt);
\filldraw[black] (3.7,3.7) circle (2.5pt);
\filldraw[black] (-0.9,5.1) circle (2.5pt);
\filldraw[black] (-3.8,5.2) circle (2.5pt);
\draw[decorate,very thick,decoration={zigzag,segment length=1.5mm,amplitude=0.5mm}] (-0.9,5.1) to[out=160,in=-10] (-3.8,5.2);
%
%
\node at (1.5,0.3) {\footnotesize bath};
\node at (-1.5,0.3) {\footnotesize AdS};
\node at (1.7,3.5) {\footnotesize $R$};
\node at (-2.1,5.7) {\footnotesize $A$};
%
\node at (0.2,4.75) {\footnotesize $t$};
\node at (-0.6,4.8) {\footnotesize $p_h$};
\node at (-4.3,5.2) {\footnotesize $p_0$};
%
%
\end{tikzpicture}
\caption{\footnotesize The Hawking radiation $R$ of an old black hole, emitted in the interval $\langle 0,t\rangle$. The modes $A$ behind the horizon are contained in an interval $[p_0,p_h]$ where $p_h$ is on the horizon.}
\label{fig5} 
\end{center}
\end{figure}

In order to put some flesh on the bones, let us consider the radiation $R$ emitted in the interval $\langle 0,t\rangle$ past the Page time $t>t_\text{Page}$ and a large subset of the modes behind the horizon $A$: see figure \ref{fig5}. The  argument is similar to that in \cite{Harlow:2014yka}. Past the Page time, where $S_R$ is decreasing, we can expect---and verify later---that $S_A>S_R$ and so the triangle inequality of Araki and Lieb \eqref{tie} implies 
\EQ{
S_{AR}\geq S_A-S_R\ .
\label{huy}
}
This can be written as
\EQ{
I_{A,R}\leq 2S_R\ .
\label{yiw}
}
So the fact that the late modes are entangled with the early modes limits the mutual information of the radiation with the modes inside the horizon.

Now in order for the inequality \eqref{yiw} to have any teeth, we have to assume that the UV cut is not so large so that $S_\text{BH}\gg |c\log\varepsilon_\text{UV}|$ and hence the cut off terms are small compared with the ${\mathscr O}(k^{-1})$ contributions to the entropy. Making this assumption, let us calculate the mutual information $I_{A,R}$. The interval $A$ is the region behind the horizon between points $p_0$ and $p_h$, where $p_0$ is some point deep inside the black hole and $p_h$ is a point on the horizon, i.e.~with coordinate $w_h^-=0$. The interval in the bath is $[0,t]$ between points $p_1$ on the boundary, with coordinate $w^-=-\hat f(t)^{-1}$, and $p_2$ far into the bath.

The first issue to settle is whether the interval behind the horizon $A$, or $AR$, can have an island saddle itself. There is no reason, a priori, why they cannot. We do not have a general proof that there are no island saddles that can contribute, but suppose we try to have a QES $p_{\hat1}$ somewhere near the horizon. The extremization will yield a pair of equations dominated by the interaction between $p_h$ and the QES:\footnote{We ignore the distinction between $\pmb{\omega}_{\hat 1}^+$ and $w_{\hat 1}^+$ which is sub-leading in the exponent}
\EQ{
\frac{2\pi}{\beta k}w_{\hat 1}^+-\frac1{w_{\hat 1}^--w_h^-} +\cdots=0\ ,\qquad\frac{2\pi}{\beta k}w_{\hat1}^--\frac1{4w_{\hat1}^+}+\cdots=0\ .
\label{quf5}
}
This means that
\EQ{
w^+_{\hat1}=-\frac{3\beta k}{8\pi}\cdot\frac1{w^-_h}\ ,\qquad w^-_{\hat1}=-\frac34w^-_h
}
and so as $p_h$ approaches the horizon, $w^-_h\to0$, the QES to pushed off to an unphysical $w^+_{\hat1}\to\infty$, $w^-_{\hat 1}\to0$. While this is not a proof that an island cannot dominate, it is suggestive and we will assume that it is true for $A$ and $AR$.

If $R$ were in its no-island saddle, then the mutual information $I_{A,R}$ would come from the cross terms between the points inside the black hole and those in the bath. However, $R$ {\it is\/} in its island-$(1)$ saddle and therefore, 
\EQ{
I_{A,R}\thicksim-\frac c6\log\frac{\sigma_{h1}\sigma_{02}}{\sigma_{01}\sigma_{h2}}+S_{R,\text{island}(1)}-S_{R,\text{no island}}\ ,
\label{pop}
}
where the $\sigma$'s are expressed in the $(y^+,w^-)$ frame. We have $S_{R,\text{island}(1)}=S_\text{BH}e^{-kt/2}$ and $S_{R,\text{no island}}=2S_\text{BH}(1-e^{-kt/2})$. The cross terms are dominated by the $\alpha_{h1}$ term, and in particular the piece coming from the difference of the $w^-$ coordinates which is becoming small at late times; hence
\EQ{
I_{A,R}&\thicksim-\frac c6\log(\hat f_1^{-1}-0)-S_\text{BH}\big(2-3e^{-kt/2}\big)\\
&\to 4S_\text{SB}{\cal S}(t)-S_\text{BH}\big(2-3e^{-kt/2}\big)=S_\text{BH}\big(2-e^{-kt/2}\big)\ .
\label{vup}
}
This is in contradiction with the expectation \eqref{yiw} precisely when $t>t_\text{Page}$, and implies some radical departure from the idea that the inside of the black hole is a separate subsystem.

We can put our finger on exactly where the monogamy argument goes wrong. From  \eqref{vup}, since $S_A=S_{AR}+I_{A,R}-S_R$, this clearly indicates a breakdown of the Araki-Lieb triangle inequality:
\EQ{
S_A-S_R= S_{AR}+S_\text{BH}\big(2-3e^{-kt/2}\big)\nless S_{AR}\ ,
}
precisely for $t>t_\text{Page}$. Intuitively, what is happening is that $S_R$ is dominated by a saddle that includes the island in the interior of the black hole, but $S_{AR}$ has no island because it already includes the interior region. So the subsystems $R$ and $A$ actually overlap because $R$ has an island, and the triangle inequality only applies when the two subsystems are distinct. Note that the Araki-Lieb inequality was also identified as being at risk in the related scenario in \cite{Almheiri:2019psf}.

These considerations lend weight to the so-called ``$A=R_B$" hypothesis (e.g.~see the review \cite{Harlow:2014yka}) which identifies the Hilbert space of modes in the early radiation $R_B$, that are entangled with late modes $B$, with the Hilbert space of modes behind the horizon $A$. This leads to the astonishing conclusion that the modes inside an old black hole are actually modes of the Hawking radiation emitted earlier.

\section{Recovering information}\label{s7}

In this section, we consider the effect of an additional shockwave inserted on the boundary at a later time. The shockwave has negligible energy, so there is no significant back reaction,  but carries entanglement entropy between its out- and in-going components. Since it is information carrying, as is the fashion \cite{Hayden:2007cs}, we will refer to the in-going shockwave as the ``diary" and the out-going one as its ``purifier".

Before we turn to the gravitational setting, let us consider the effect in a CFT in Minkowski space. Specifically, let us consider the effect on the mutual information of 2 intervals $D$ and $P$. As a component of the shockwave, e.g.~the diary, enters an interval, say $D$, the entropies $S_D$ and $S_{DP}$ jump by $S_\text{diary}$ so the mutual information stays the same. If $D$ contains the diary and $P$ the purifier, both $S_D$ and $S_P$ jump by $S_\text{diary}$ but $S_{DP}$ stays the same. So the mutual information $I_{D,P}$ jumps by $2S_\text{diary}$. This makes intuitive sense, if $D$ and $P$ contain the diary and the purifier, respectively, the mutual information increases by $2S_\text{diary}$.

\begin{figure}[ht]
\begin{center}
\begin{tikzpicture} [scale=0.7]
\filldraw[color=black!20,draw=none] (-5,2.8) -- (-5,1.2) -- (-0.5,5.7) -- (-1.3,6.5) -- (-5,2.8); 
\filldraw[color=black!35,draw=none] (-0.8,0.8) -- (3.2,4.8) -- (1.95,6.05) -- (-2.05,2.05) -- (-0.8,0.8); 
\filldraw[color=black!35,draw=none] (0,0) -- (4,4) -- (3.5,4.5) -- (-0.5,0.5) -- (0,0); 
\draw[thick,dashed] (0,0) to[out=90,in=-105] (1,7);
\draw[thick] (1,7) -- (5,3);
\draw[thick,dash dot] (5,0)  -- (-5,0);
\draw[double,decorate,very thick,->,decoration={snake,amplitude=0.03cm}] (0,0) -- (-5,5);
\draw[double,decorate,very thick,->,decoration={snake,amplitude=0.03cm}] (0,0) -- (4,4);
\draw[decorate,very thick,decoration={zigzag,segment length=1.5mm,amplitude=0.5mm}] (-5,2.7) to[out=43,in=-155] (-0.5,5.7);
\filldraw[black] (-0.5,5.7) circle (2.5pt);
\draw[decorate,very thick,-triangle 60,decoration={snake,amplitude=0.03cm}] (0.1,1.4) -- (3.35,4.65);
\draw[decorate,very thick,-triangle 60,decoration={snake,amplitude=0.03cm}] (0.1,1.4) -- (-4.1,5.7);
%
%
\draw[decorate,very thick,decoration={zigzag,segment length=1.5mm,amplitude=0.5mm}] (0.5,4.6) to[out=-5,in=145] (3.35,3.35);
\filldraw[black] (0.5,4.6) circle (2.5pt);
%
\draw[thick] (-5,3) -- (-4,4);
\draw[thick] (-3,3) -- (1,7);
%
\filldraw[black] (3.35,3.35) circle (2.5pt);
\filldraw[black] (2.8,3.75) circle (2.5pt);
\filldraw[black] (2.35,4.05) circle (2.5pt);
\node at (1.5,0.3) {\footnotesize bath};
\node at (-1.5,0.3) {\footnotesize AdS};
\node[rotate=-45] at (-1.45,3.5) {\footnotesize diary};
\node[rotate=45] at (1.4,3.2) {\footnotesize purifier};
\node at (2,5) {\footnotesize $D$};
\node at (3.7,5) {\footnotesize $P$};
\node[rotate=45] at (-4.3,2.7) {\footnotesize island};
\node at (0.2,4.7) {\footnotesize $t$};
\node at (2.8,1.4) (a1) {\footnotesize $t_0$};
\draw[thick,dotted] (a1) -- (0.1,1.4);
\end{tikzpicture}
\caption{\footnotesize In this scenario, the Hawking radiation is collected in interval $D$ from the beginning of the evaporation, excluding a small interval $P$ that is used to collect the purifier of the diary. Also shown is the island of $D$ that dominates when $t$ is large enough. The information is returned in $D$ when $t$ is a little larger than the Page time determined by the the diary's entropy.}
\label{fig6} 
\end{center}
\end{figure}
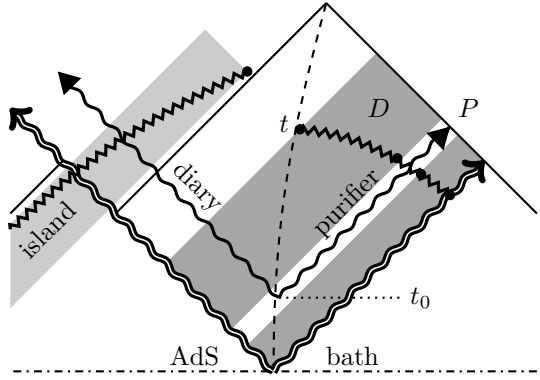

\subsection{Black hole as a mirror scenario}

Within the JT gravity framework, there are several scenarios we could discuss. The simplest is the one that matches the thought experiments of Hayden and Preskill \cite{Hayden:2007cs}, where a diary, i.e.~a local entanglement carrying quench, is dropped into the black hole at some point on boundary at time $t_0>0$: see figure \ref{fig6}. All the Hawking radiation is collected up to some time $t$ and the question is how large does $t$ have to be to recover the information of diary. To this end, we reserve a narrow interval in bath $P=[t-t_0-\delta,t-t_0+\delta]$, designed to pick up the purifier at time $t$, where $\delta$ is small. The complement $D=[0,t-t_0-\delta]\cup[t-t_0+\delta,t]$, is then the Hawking radiation collected in this context. The information has returned at time $t$ if
\EQ{
I_{D,P}(t)\approx2S_\text{diary}\ .
\label{upp}
} 
In order for the information to be returned, requires that the islands of $D$ and $DP$ contain the diary, whereas $P$ must not have an island.\footnote{There is a subtlety here because $D$ consists of 2 intervals, albeit ones separated by a small amount, and so has two island saddles: in the notation of \eqref{twointerval} island-$(1)$ and island-$(3)$. In this case, the saddle $(3)$ is always sub-dominant 
. Then the non-diary parts of $S_D$ and $S_{DP}$ for the no-island and island saddles are approximately the same.} This latter requirement is guaranteed since $P$ is a narrow interval and, as long as the entropy of the diary is not too large compared with $S_\text{BH}$, the no-island saddle always dominates:
\EQ{
S_{P,\text{no island}}&=2S_\text{BH}e^{-kt_0/2}(e^{k\delta/2}-e^{-k\delta/2})+S_\text{diary}\approx S_\text{diary}\ ,\\[5pt]
S_{P,\text{island}(1)}&=S_\text{BH}\big(2-e^{-kt_0/2}(2e^{k\delta/2}-e^{-k\delta/2})\big)\approx S_\text{BH}(2-e^{-kt_0/2})\ ,
}
In the above, note that the no-island saddle is shifted by $S_\text{diary}$ because $P$ contains the purifier but the island contains the diary and so the island saddle is not shifted. We are assuming that diary entropy is not too large compared with the black hole so $\eta=S_\text{diary}/S_\text{BH}\ll1$.

So in order for the information to be recovered, $D$ and $DP$ must be dominated by their island-(1) saddles and the diary must be contained within the islands, so that
\EQ{
S_{DP}\approx S_D-S_\text{diary}\ ,
}
from which \eqref{upp} follows.

Let us consider $D$. Although this consists of 2 intervals, the non-diary part of the entropy is approximately that of a single interval $[0,t]$, since $\delta$ is small. Of course $DP$ is the whole interval $[0,t]$. The difference is that when when the diary is in the island of $D$ the island entropy is shifted {\em up} by $S_\text{diary}$, whereas for $DP$ the opposite is true, the no-island entropy is shifted by $S_\text{diary}$ since it contains the purifier. So the island  for $DP$ appears at $t_-$, {\em before} that of $D$ at time $t_+$, where
\EQ{
2-2e^{-kt_\mp/2}=e^{-kt_\mp/2}\mp\eta\quad\implies\quad t_\mp\approx t_\text{page}\mp\frac{\eta}k\ .
\label{wul}
}
However, we have to check that the diary is in the island, that is the $w^+$ coordinate of the diary is less than that of the QES $p_{\hat2}$ of $DP$, or $D$, at time $t$ when the radiation is collected, i.e.
\EQ{
\hat f(t_0)<\frac{3\beta k}{8\pi}\hat f(t)\,e^{kt/2}\ .
}
The pre-factor here, leads to a delay $t>t_0+\Delta t_\text{scr.}$,
\EQ{
\Delta t_\text{scr.}\approx\frac\beta{2\pi}e^{kt/2}\log\left(\frac{8\pi}{3\beta k} e^{-kt/2}\right)\ ,
}
identified as the scrambling time of the black hole, however, this is a small effect when the leading time scales are of order $k^{-1}$ and so, to leading order, we ignore it. Collecting all this together yields the mutual information, which increases in a piece-wise continuous fashion as
\EQ{
I_{D,P}(t)=\begin{cases} 0 & t\leq t_-\ ,\\ S_\text{BH}\big(2-3e^{-kt/2}\big)+S_\text{diary} & t_-\leq t\leq t_+\ ,\\ 2S_\text{diary} & t_+\leq t\ .\end{cases}
}

So the following picture emerges that is entirely consistent with the analysis of Hayden and Preskill \cite{Hayden:2007cs}. If we collect all the Hawking radiation emitted from $t=0$, then if the diary is thrown in before the time $t_-$, a little earlier than the Page time, then the information will be returned a little later than the Page time at $t_+$. From \eqref{wul}, note that the additional time beyond $t_\text{Page}$ is proportional to the diary's entropy. If the diary is thrown in after $t_+$, then it is returned almost immediately after a short delay given by the instantaneous scrambling time of the black hole $\Delta t_\text{scr.}$ at time $t_0$. This is the origin of the ``black hole as a mirror" metaphor. 

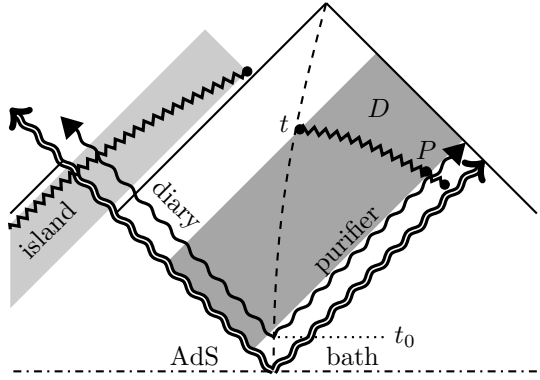
\begin{figure}[ht]
\begin{center}
\begin{tikzpicture} [scale=0.7]
\filldraw[color=black!20,draw=none] (-5,2.8) -- (-5,1.2) -- (-0.5,5.7) -- (-1.3,6.5) -- (-5,2.8); 
\filldraw[color=black!35,draw=none] (-0.4,0.4) -- (3.6,4.4) -- (1.95,6.05) -- (-2.05,2.05) -- (-0.4,0.4); 
\draw[thick,dashed] (0,0) to[out=90,in=-105] (1,7);
\draw[thick] (1,7) -- (5,3);
\draw[thick,dash dot] (5,0)  -- (-5,0);
\draw[double,decorate,very thick,->,decoration={snake,amplitude=0.03cm}] (0,0) -- (-5,5);
\draw[double,decorate,very thick,->,decoration={snake,amplitude=0.03cm}] (0,0) -- (4,4);
\draw[decorate,very thick,decoration={zigzag,segment length=1.5mm,amplitude=0.5mm}] (-5,2.7) to[out=43,in=-155] (-0.5,5.7);
\filldraw[black] (-0.5,5.7) circle (2.5pt);
\draw[decorate,very thick,-triangle 60,decoration={snake,amplitude=0.03cm}] (0,0.65) -- (3.65,4.35);
\draw[decorate,very thick,-triangle 60,decoration={snake,amplitude=0.03cm}] (0,0.65) -- (-4.05,4.85);
%
\draw[decorate,very thick,decoration={zigzag,segment length=1.5mm,amplitude=0.5mm}] (0.5,4.6) to[out=-5,in=150] (3.25,3.55);
\filldraw[black] (0.5,4.6) circle (2.5pt);
%
\draw[thick] (-5,3) -- (-4,4);
\draw[thick] (-3,3) -- (1,7);
%
\filldraw[black] (3.25,3.55) circle (2.5pt);
\filldraw[black] (2.9,3.8) circle (2.5pt);
\node at (1.5,0.3) {\footnotesize bath};
\node at (-1.5,0.3) {\footnotesize AdS};
\node[rotate=-45] at (-1.85,3.15) {\footnotesize diary};
\node[rotate=45] at (1.4,2.5) {\footnotesize purifier};
\node at (2,5) {\footnotesize $D$};
\node at (2.9,4.2) {\footnotesize $P$};
\node[rotate=45] at (-4.3,2.7) {\footnotesize island};
\node at (0.2,4.7) {\footnotesize $t$};
\node at (2.5,0.65) (a1) {\footnotesize $t_0$};
\draw[thick,dotted] (a1) -- (0,0.65);
\end{tikzpicture}
\caption{\footnotesize In this scenario, the Hawking radiation is collected after the diary is thrown into the black hole at $t_0$. The information is returned in $D$ if $t_0$ is not too large and $t$ is sufficiently late.}
\label{fig40} 
\end{center}
\end{figure}

\subsection{Recovery after the fact}

There are variations of this scenario we can analyse. Suppose we throw the diary in, but only then start to collect the Hawking radiation in a temporal interval $\langle t_0,t\rangle$, as shown in figure \ref{fig40}. In this case, we collect the Hawking radiation in a spatial interval $D=[0,t-t_0]$ at time $t$. Now the conditions \eqref{wul} that $DP$ and $D$ have an island at $t=t_\mp$, respectively, are modified to
\EQ{
&2e^{-kt_0/2}-2e^{-kt_\mp/2}=e^{-kt_\mp/2}+2-2e^{-kt_0/2}\mp\eta\\ &\implies\quad t_\mp=t_0+\frac 2k\log\frac3{4-(2\mp\eta)e^{kt_0/2}}\ .
}
We know that in order to recover the information at time $t$, i.e.~$I_{D,P}(t)\approx2S_\text{diary}$, requires $t\geq t_+$. It follows that the diary cannot be thrown in too late
\EQ{
t_0<\frac2k\log\frac4{2+\eta}\ .
\label{pep}
}
In addition, the minimum collection time $t_+-t_0$ becomes longer for later $t_0$:
\EQ{
t_+-t_0=\frac 2k\log\frac3{4-(2+\eta)e^{kt_0/2}}\ .
}
Note that if the diary is thrown in close to, or later than, the bound \eqref{pep}, $t_+$ becomes too large or diverges and our approximations will break down. One expects that the exact analysis will show that the information does come out eventually, as the black hole evaporates back to the extremal black hole.

\section{Behind the horizon in the bath}\label{s8}

The entropy calculations show that for an old black hole the physics behind the horizon is subtly encoded in the state of the bath. This extends to more refined quantities like the entropy of an interval in the bath in an excited state of the CFT created by the local operator insertion, as our story of recovering information in section \ref{s7} shows. In this section, we consider this in a little more detail. The question we ask, is how does the entropy of an interval of radiation in the bath respond to an operator insertion behind the horizon?

First of all, in the non-gravitational setting, the response of an interval to an operator insertion has been well studied, e.g.~\cite{He:2014mwa,David:2016pzn}. The excited state is ${\cal O}(y^\pm_0+i\varepsilon)\ket{0}$, defined as in \eqref{lic} with a small imaginary shift in the insertion point in order to ensure that the state is normalized.
If the interval is $A=[b_1,b_2]$ and the operator is inserted at point $y^\pm_0$, then the entropy $\Delta S_A(t)$, the difference of the entropy with and in the absence of the operator insertion, responds in a causal way. This means that $\Delta S_A(t)$ becomes non-trivial approximately when either $t-b_2\leq y^-_0\leq t-b_1$ or $t+b_1\leq y^+_0\leq t+b_2$. 

The intuitive picture is that the operator insertion creates a left- and a right-moving shockwave and the entropy of the interval responds as either of these shockwaves moves through the interval. The leading order effect for small $\varepsilon$, at $\mathscr{O}(\varepsilon^0)$, is that $\Delta S_A(t)$ jumps by $\log d_{\cal O}$, where $d_{\cal O}$ is the {\it quantum dimension\/} of the operator---not to be confused with its scaling dimension---when either of the shockwave moves into the interval. The quantity $\log d_{\cal O}$ is a measure of the entanglement of the operator ${\cal O}$ between the left- and right-moving shockwaves. In section \ref{s7}, we called $\log d_{\cal O}$ the entropy of the diary. There are interesting universal corrections in an expansion in the small quantity $\varepsilon$ \cite{David:2016pzn}.

Given how the entropy of a subregion behaves in the presence of an operator insertion, it seems obvious that, when an interval in the bath is in an island saddle, operator insertions behind the horizon that pass through the island will be witnessed by the interval in the bath. This seems to be a prima facie violation of locality that is, in principle, measurable in the bath in the sense that we describe in the discussion section \ref{s1.1}.

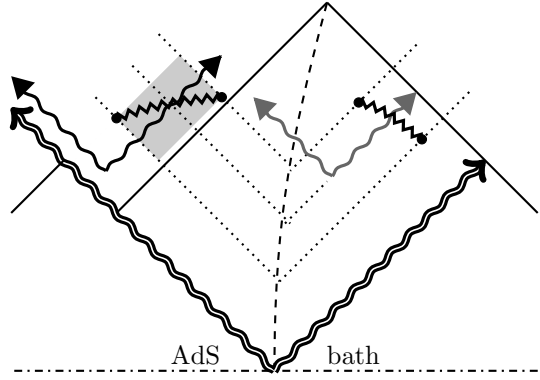
\begin{figure}[ht]
\begin{center}
\begin{tikzpicture} [scale=0.7]
\filldraw[color=black!20,draw=none] (-3,4.8) -- (-2.2,4) -- (-1,5.2) -- (-1.8,6) -- (-5,2.8); 
\draw[thick,dotted] (-3.4,5.2) -- (0.1,1.7);
\draw[thick,dotted] (-2.2,6.4) -- (0.3,3.9);
\draw[thick,dotted] (3.7,5.3) -- (0.1,1.7);
\draw[thick,dotted] (2.7,6.2) -- (0.3,3.8);
\draw[thick,dotted] (-2.7,5.8) -- (0.3,2.8);
\draw[thick,dotted]  (0.2,2.8) -- (1,3.6);
\draw[thick,dashed] (0,0) to[out=90,in=-105] (1,7);
\draw[thick] (1,7) -- (5,3);
\draw[thick,dash dot] (5,0)  -- (-5,0);
\draw[double,decorate,very thick,->,decoration={snake,amplitude=0.03cm}] (0,0) -- (-5,5);
\draw[double,decorate,very thick,->,decoration={snake,amplitude=0.03cm}] (0,0) -- (4,4);
\draw[decorate,very thick,-triangle 60,decoration={snake,amplitude=0.03cm}] (-3.2,3.8) -- (-5,5.6);
\draw[decorate,very thick,-triangle 60,decoration={snake,amplitude=0.03cm}] (-3.2,3.8) -- (-1,6);
\draw[decorate,very thick,black!60,-triangle 60,decoration={snake,amplitude=0.03cm}] (1.1,3.7) -- (2.7,5.3);
\draw[decorate,very thick,black!60,-triangle 60,decoration={snake,amplitude=0.03cm}] (1.1,3.7) -- (-0.4,5.2);
\draw[decorate,very thick,decoration={zigzag,segment length=1.5mm,amplitude=0.5mm}] (-3,4.8) to[out=20,in=-175] (-1,5.2);
\filldraw[black] (-1,5.2) circle (2.5pt);
\filldraw[black] (-3,4.8) circle (2.5pt);
\draw[decorate,very thick,decoration={zigzag,segment length=1.5mm,amplitude=0.5mm}] (1.6,5.1) to[out=-10,in=145] (2.8,4.4);
\filldraw[black] (1.6,5.1) circle (2.5pt);
\filldraw[black] (2.8,4.4) circle (2.5pt);
%
%
%
\draw[thick] (-5,3) -- (-4,4);
\draw[thick] (-3,3) -- (1,7);
\node at (1.5,0.3) {\footnotesize bath};
\node at (-1.5,0.3) {\footnotesize AdS};
\end{tikzpicture}
\caption{\footnotesize An operator insertion made behind the horizon generates a pair of shockwaves. Here, at a certain time, the right moving one is in the island of the interval in the bath, the shaded region, and creates an effect that is indistinguishable from an operator insertion made in the bath as shown.}
\label{fig15} 
\end{center}
\end{figure}

Let us consider the effect in more detail. Let us take an interval in the bath $A=[b_1,b_2]$. At late times, and certainly $t>b_2$, the entropy of the interval involves a competition between the no-island and the island-$(21)$ saddles:
\EQ{
S_\text{no island}&=2S_\text{BH}e^{-kt/2}\big(e^{kb_2/2}-e^{kb_1/2}\big)\ ,\\[5pt]
S_{\text{island}(21)}&=S_\text{BH}e^{-kt/2}\big(e^{kb_1/2}+e^{kb_2/2}\big)\ .
}
The island-$(21)$ saddle dominates when $e^{kb_2/2}>3e^{kb_1/2}$. This saddle illustrated in figures \ref{fig7} and \ref{fig15}.

Suppose the operator insertion behind the horizon as shown in figure \ref{fig15}, at $w^\pm_0$. In this case, the right-moving shockwave lies in the island if
\EQ{
\frac1{3\hat f(t-b_2)}\leq w_0^-\leq\frac1{3\hat f(t-b_1)}\ .
}
So the effect is indistinguishable to a scenario where the operator insertion in made in the bath at a point with coordinate $y_0^-=\hat f^{-1}(1/(3w_0^-))$.

\section{Discussion}\label{s9}

In this work we set up the calculation of  von Neumann entropies for an arbitrary number of subsystems within the Hawking radiation bath of an evaporating black hole in JT gravity produced by a shockwave injected via a local quench. The primary goal in this analysis was to display the correlations 
within early and late Hawking modes, and to flesh out the ``$A {=} R_B$" scenario within this framework. The slow evaporation limit $k\ll1$ with $kt$ fixed, in conjunction with  large interval sizes $\sim {\mathscr O}(k^{-1})$ results in a simple but rich, analytically tractable interplay of island and no-island saddles.  We focussed attention on island saddles with only two QESs since we generically expect additional QES contributions to significantly increase the entropy. Such islands are sufficient for teasing out  the refined correlations between two disjoint subsets of the Hawking modes in the bath. Whether islands with additional QESs dominate,
 likely depends on the number and size of separations between the intervals in the bath, and it would be very interesting to understand  the conditions under which such saddles could become relevant, and how multi-QES islands may compete with each other.  In the bath CFT, R\'enyi/entanglement entropies can be understood as correlators of twist fields. The operator product of twist fields is dominated by the stress tensor in the short interval limit. 
 It would be interesting to understand what implications the calculation of multi-interval entropies has for correlators of local operators (such as the stress tensor) and lack of cluster property (see e.g. \cite{Blommaert:2020seb}) in the presence of replica wormhole (island saddle) contributions.

\vspace{0.5cm}
\begin{center}{\it Acknowledgments}\end{center}
\vspace{0.2cm}
 We would like to thank Justin David, Carlos N\'u\~nez and Dan Thompson for discussions. TJH and SPK acknowledge support from STFC grant ST/P00055X/1.

\end{document}